\documentclass[sigconf]{acmart}

\usepackage{subcaption}  
\usepackage{multirow}  
\usepackage{xspace}

\AtBeginDocument{%
  \providecommand\BibTeX{{%
    \normalfont B\kern-0.5em{\scshape i\kern-0.25em b}\kern-0.8em\TeX}}}

\setcopyright{iw3c2w3}
\copyrightyear{2021}
\acmYear{2021}
\acmDOI{10.1145/3442381.3449947}
\acmConference[WWW '21]{Proceedings of the Web Conference 2021}{April 19--23, 2021}{Ljubljana, Slovenia}
\acmBooktitle{Proceedings of the Web Conference 2021 (WWW '21), April 19--23, 2021,
Ljubljana, Slovenia}
\acmPrice{}
\acmISBN{978-1-4503-8312-7/21/04}

\begin{document}

\title{Learning Heterogeneous Temporal Patterns of User Preference for Timely Recommendation}

\author{Junsu Cho}
\affiliation{%
  \institution{Pohang University of Science and Technology}
  \city{Pohang}
  \country{South Korea}}
\email{junsu7463@postech.ac.kr}

\author{Dongmin Hyun}
\affiliation{%
  \institution{Pohang University of Science and Technology}
  \city{Pohang}
  \country{South Korea}}
\email{dm.hyun@postech.ac.kr}

\author{SeongKu Kang}
\affiliation{%
  \institution{Pohang University of Science and Technology}
  \city{Pohang}
  \country{South Korea}}
\email{seongku@postech.ac.kr}

\author{Hwanjo Yu}
\authornote{Corresponding Author}
\affiliation{%
  \institution{Pohang University of Science and Technology}
  \city{Pohang}
  \country{South Korea}}
\email{hwanjoyu@postech.ac.kr}


\begin{abstract}
Recommender systems have achieved great success in modeling user's preferences on items and predicting the next item the user would consume. Recently, there have been many efforts to utilize time information of users' interactions with items to capture inherent temporal patterns of user behaviors and offer timely recommendations at a given time. Existing studies regard the time information as a single type of feature and focus on how to associate it with user preferences on items. However, we argue they are insufficient for fully learning the time information because the temporal patterns of user preference are usually heterogeneous. A user's preference for a particular item may 1) increase periodically or 2) evolve over time under the influence of significant recent events, and each of these two kinds of temporal pattern appears with some unique characteristics. In this paper, we first define the unique characteristics of the two kinds of temporal pattern of user preference that should be considered in time-aware recommender systems. Then we propose a novel recommender system for timely recommendations, called TimelyRec, which jointly learns the heterogeneous temporal patterns of user preference considering all of the defined characteristics. In TimelyRec, a cascade of two encoders captures the temporal patterns of user preference using a proposed attention module for each encoder. Moreover, we introduce an evaluation scenario that evaluates the performance on predicting an interesting item and when to recommend the item simultaneously in top-$K$ recommendation (i.e., item-timing recommendation). Our extensive experiments on a scenario for item recommendation and the proposed scenario for item-timing recommendation on real-world datasets demonstrate the superiority of TimelyRec and the proposed attention modules. 
\end{abstract}




\maketitle

\section{Introduction}
 Recommender system has been actively researched and successfully applied to many real-world services by modeling users' interests on items. To increase the amount of information to learn and provide more personalized recommendations, recommender systems often utilize side information in addition to the interactions (e.g., click or buy) between users and items, such as user profile \cite{userprofile, userprofile2}, item category \cite{itemcategory, itemcategory2}, or \textit{time information} \cite{ntf, concars, timelstm, slirec}. Time information, which is information about when users left their records of interactions with items, implies various time-related patterns of user behaviors \cite{temporal1, temporal2}, while it can be obtained without any additional request to the user. Recently, many recommender systems have been studying how to utilize the time information to effectively capture behavioral patterns of the users and provide timely recommendations to them \cite{ntf, concars, timelstm, slirec, tisasrec}. 
 
 \begin{figure}[t]
    \centering
    \includegraphics[scale=0.32]{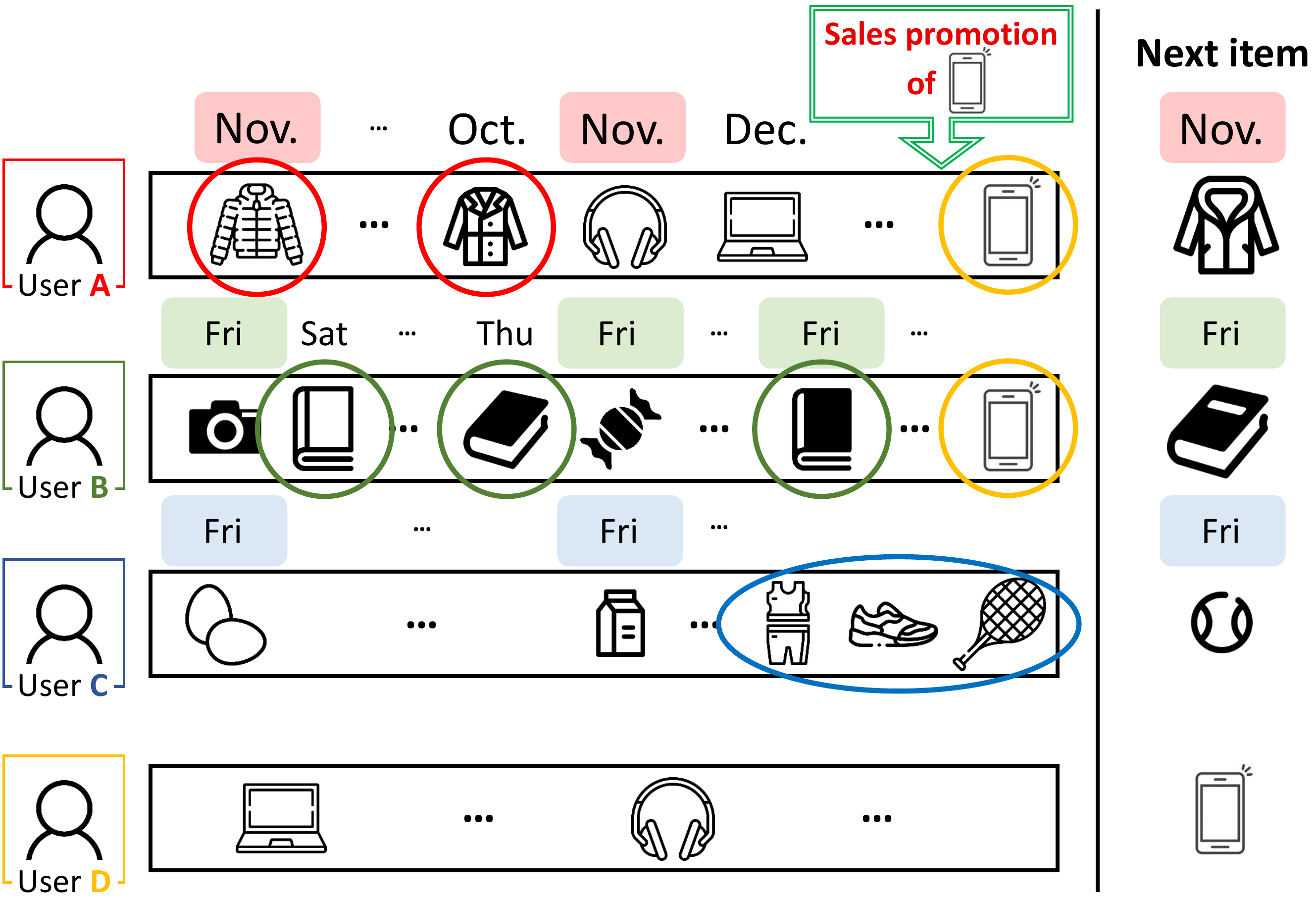}
    \caption{An example of users' purchase sequence. The circles of each color indicate items related to the next item of the corresponding user.}
    \label{fig:intro}
 \end{figure}
 
 Recommender systems comprehending the inherent temporal patterns of user behaviors can accomplish various time-related tasks. First, if a recommender system understands the users' preference over time, it can provide users with timely recommendations when requested by the users (i.e., \emph{item recommendation}). For example, a music streaming service can recommend a music that the user may like according to the time the user accesses. Second, a time-aware recommender system can predict an interesting item and when the user is likely to consume the item simultaneously, and actively recommend the item at that time (i.e., \emph{item-timing recommendation}). For example, a mobile application can provide recommendations via push notifications whenever a user is expected to have interest in a particular item. Also, an e-commerce platform may predict in advance when an item will be in demand and prepare for the surge in demand. 

 Existing methods consider the time information as a single type of feature, and focus on associating it with user preferences on items \cite{ntf, concars, timelstm, slirec, tisasrec}. However, we argue it is insufficient for fully learning the time information in that the temporal patterns of user preference are usually heterogeneous. In other words, a user's preference for a particular item may either increase periodically (\emph{periodic pattern}), or evolve over time under the influence of significant recent events (\emph{evolving pattern}) \cite{concars}, and we also argue that each of these two kinds of temporal pattern appears with some unique characteristics that should be carefully considered.
 
 In this regard, we first define the unique characteristics of the two kinds of temporal patterns of user preferences that should be considered in time-aware recommender systems (Fig. \ref{fig:intro}). First, the \emph{periodic pattern} of user preferences has three unique characteristics: 1) it has multiple granularity of period (User A vs. User B). For example, a user may like to listen to a ballad song around midnight (hour), watch an action movie on Friday (day of the week), buy an expensive item on payday (date), or buy a winter coat in November (month). 2) The periodic pattern is often slightly irregular (User A and User B). For example, the user may sometimes buy a winter coat in October or December, not in November. 3) Each user's behavioral pattern is personalized, because everyone has their own lifestyle (User B vs. User C). Second, the \emph{evolving pattern} of user preferences leaves us with two kinds of significant events to observe closely, which can affect the user's subsequent preferences: 1) the user's recent interactions  such as purchase of related items (User C) or 2) events that create temporal trends of items such as promotions (User D).

 In this paper, we propose a novel recommender system for timely recommendations, called TimelyRec, which jointly considers all the unique characteristics of the heterogeneous temporal patterns. 
 TimelyRec consists of a cascade of two main components, each of which encodes the information of a kind of temporal pattern into a latent representation.
 Specifically, \emph{Multi-Aspect Time Encoder} (MATE) (Sec. \ref{multi_aspect_time_encoder}) encodes the multi-aspect information about periodic patterns of the given timestamp. In other words, considering the characteristics of periodic pattern, MATE personalizes each of the multiple granularity of period information (i.e., month, day of the week, date, and hour), captures the potential irregularity in each of them, and adaptively adopts each of them depending on the user. To this end, we propose a novel attention module called \emph{gradual attention} which effectively captures the irregularity of periodic pattern. Our another encoder, \emph{Time-Aware History Encoder} (TAHE) (Sec. \ref{time_aware_history_encoder}), learns the evolving patterns of user preference. TAHE utilizes the output of MATE to encode the user's recent interactions. The key idea is to give more weight to the interactions made at a time that has similar temporal pattern to the target time. We implement this idea via an attention module that we propose named \emph{time-based attention}. Moreover, we adopt sinusoidal positional encoding \cite{sinusoidal} to represent a temporal position of each interaction and capture the trend of items when they were consumed.

 The main contributions of this paper are summarized as follows:
 \begin{itemize}
     \item We define various unique characteristics in the periodic pattern and the evolving pattern of user preferences that should be considered in recommender systems that learn the time information.
     \item We propose a novel time-aware recommender system named TimelyRec, which jointly takes into account all the characteristics of the heterogeneous temporal patterns, and a couple of attention modules for effectively capturing some unique characteristics of temporal patterns.
     \item In addition to top-$K$ recommendation scenario for item recommendation, we introduce an evaluation scenario that evaluates the performance on item-timing recommendation in top-$K$ recommendation, which demonstrates the superiority of TimelyRec on modeling the heterogeneous temporal patterns more reliably.
     \item Our extensive experiments on real-world datasets show the superiority of TimelyRec and the proposed attention modules. TimelyRec outperforms several state-of-the-art time-aware baselines by up to 43.08\% on item recommendation, and up to 57.26\% on item-timing recommendation in terms of hit ratio. Our various analyses show that each of the characteristics we define is significant in learning the time information and improving the recommendation performance, and the attention modules in TimelyRec can provide a description of behavioral patterns of users, entrenching the superiority of the modules.
 \end{itemize}
 
\section{Related Work}
 In this section, we review several models that utilize time information, which are grouped by the type of temporal pattern that they learn from the time information. 

\begin{figure*}[t]
    \centering
    \includegraphics[clip, page=1, trim={0cm 0cm 0cm 0cm}, scale=0.45]{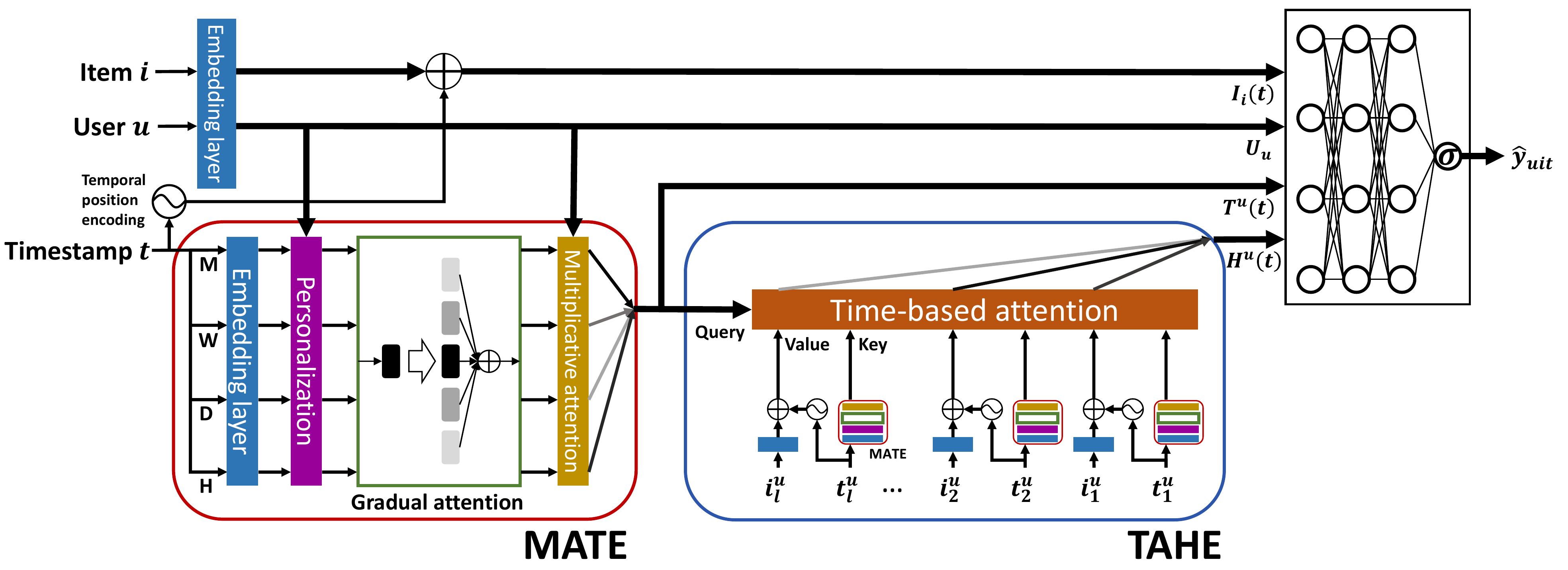}
    \caption{Model architecture of TimelyRec. M: month, W: day of the week, D: date, H: hour.}
    \label{fig:model}
\end{figure*}
 
 \subsection{Periodic Pattern}
  Some methods take into account that a user's interest in a particular item may appear periodically. Tensor Factorization (TF) \cite{parafac, pitf} is a way of modeling the interaction between the elements of input when there is another element (e.g., tag or time) besides user and item in input. PITF \cite{pitf}, which is one of TF techniques which can handle the time information, uses the dot product to compute the interaction of each pair of elements and then combines them. CoNCARS \cite{concars} represents a user using the items that the user has interacted with and the interaction times (i.e., hour), and an item in a similar way. Then CoNCARS models the non-linear interaction of the user and the item via convolutional neural networks. \cite{television} uses day of the week and hour information to recommend television programs to users, considering that users tend to watch television programs in periodic patterns in terms of day of the week and hour.
  However, they may not be able to use their knowledge when the user is interested in the item at a time slightly deviated from the regular routine in which the user normally consumes the item.
  Also they require the domain knowledge of what kind of periodic patterns are the most important for the given task.
  Moreover, they do not deal with the recent events that are hardly related to the periodic patterns but may be related to the user's current preference.

 \subsection{Evolving Pattern}
  There are also some methods that use the time information to model that a user's preference for a particular item may evolve over time under the influence of recent events. 
  NTF \cite{ntf} defines the target time as the output of LSTM \cite{lstm} taking the last few month embeddings, to capture the temporal trend of user preferences at that time. 
  However, in a practical scenario where a model is trained with past interactions and predicts future interactions, NTF must refer to untrained embedding when predicting future interactions as the model trains only the earlier embeddings during training. Therefore, the way of learning the temporal trend of items must be designed to make the information available at any time.
  Others consider the evolving pattern of the user preferences in terms of the user's recent interactions.
  TimeLSTM \cite{timelstm} uses the time intervals between consecutive interactions as well as the user's interaction history with time gates in LSTM, assuming that two consecutive interactions close to each other are more correlated to each other than two distant interactions. SLi-Rec \cite{slirec} extends TimeLSTM to use the gated mechanism and the attention mechanism to filter out unnecessary information in the user's short-term history. TiSASRec \cite{tisasrec} considers the time intervals between user's consecutive interactions in a self-attention network.
  However, they may not consider the trends of items as they do not associate the time information with items.
  Also, they do not take into account the periodic patterns of user preferences.

\section{Method}
 We propose a novel time-aware recommender system called TimelyRec (Fig. \ref{fig:model}) that jointly considers all the characteristics of the heterogeneous temporal patterns that we define. TimelyRec takes a triplet of user $u$, item $i$, and timestamp $t$ as input, and then predicts the probability that $u$ will interact with $i$ at time $t$. To this end, TimelyRec consists of a cascade of two encoders, Multi-Aspect Time Encoder (MATE) (Sec. \ref{multi_aspect_time_encoder}) that encodes the multi-aspect information about the periodic patterns of the time, and Time-Aware History Encoder (TAHE) (Sec. \ref{time_aware_history_encoder}) that models the evolving patterns of the user preference. Finally, TimelyRec considers the non-linear interactions of user, item, and all the time-related information to calculate the probability. The detailed architecture design of TimelyRec is described below.
 
\subsection{Multi-aspect Time Encoder (MATE)}
 \label{multi_aspect_time_encoder}
  To model the periodic patterns of the given timestamp $t$, MATE first takes $t$ as input and builds a personalized period information for each granularity of period. Then MATE captures potential irregularity in each kind of the personalized period information, and adaptively adopts each of them based on the task and the user.
 \subsubsection{Personalization of Period Information}
  To build a personalized period information for each granularity, we first represent the time slots for each granularity of period as embeddings. Then, because the same moment can have different meanings depending on the user, TimelyRec personalizes the time slot embeddings using user embedding. Note that we use `month' as an example of the granularity of period information in this section, but the same approach is applied to the others (i.e., day of the week, date, and hour). The personalized period information for the month of timestamp $t$ can be obtained as follows:
  \begin{equation}
      E^u_{Month}(t) = (W_{Month}U_u) \circ E_{Month}(t) 
  \end{equation}
  where $E^u_{Month}(t) \in \mathbb{R}^{d \times 1}$ is the personalized period information for the month of timestamp $t$, $W_{Month} \in \mathbb{R}^{d \times d}$ is a matrix for transforming the user embedding for personalization of month information, $U_u \in \mathbb{R}^{d \times 1}$ is the user embedding for user $u$, $\circ$ is the Hadamard product, $E_{Month}(t) \in \mathbb{R}^{d \times 1}$ is the time slot embedding for the month of timestamp $t$, and $d$ is the size of embedding. 
  
  Our personalization technique utilizes the user embedding that is already learned and used for other purpose within the model. As such, we note that our personalization method is an efficient technique in that it does not require any additional learning parameter for each user for the purpose of personalization. 
 
 \subsubsection{Capturing Irregularity of Periodic Patterns}
  Some of user behaviors that originate from periodic routines may deviate slightly from regular patterns. We note that such irregular behaviors are not likely to deviate significantly from the regular patterns. For example, the user who has purchased a winter coat every November until last year is more likely to buy a winter coat in October than in September this year.
  In this regard, we capture the irregularity of periodic patterns of user preferences by incorporating the surrounding time slots according to their distance from the target time slot, based on the personalized period information defined above.

\begin{figure}[t]
    \centering
    \includegraphics[clip, page=1, trim={0.0cm 0.0cm 0.0cm 0.0cm}, scale=0.3]{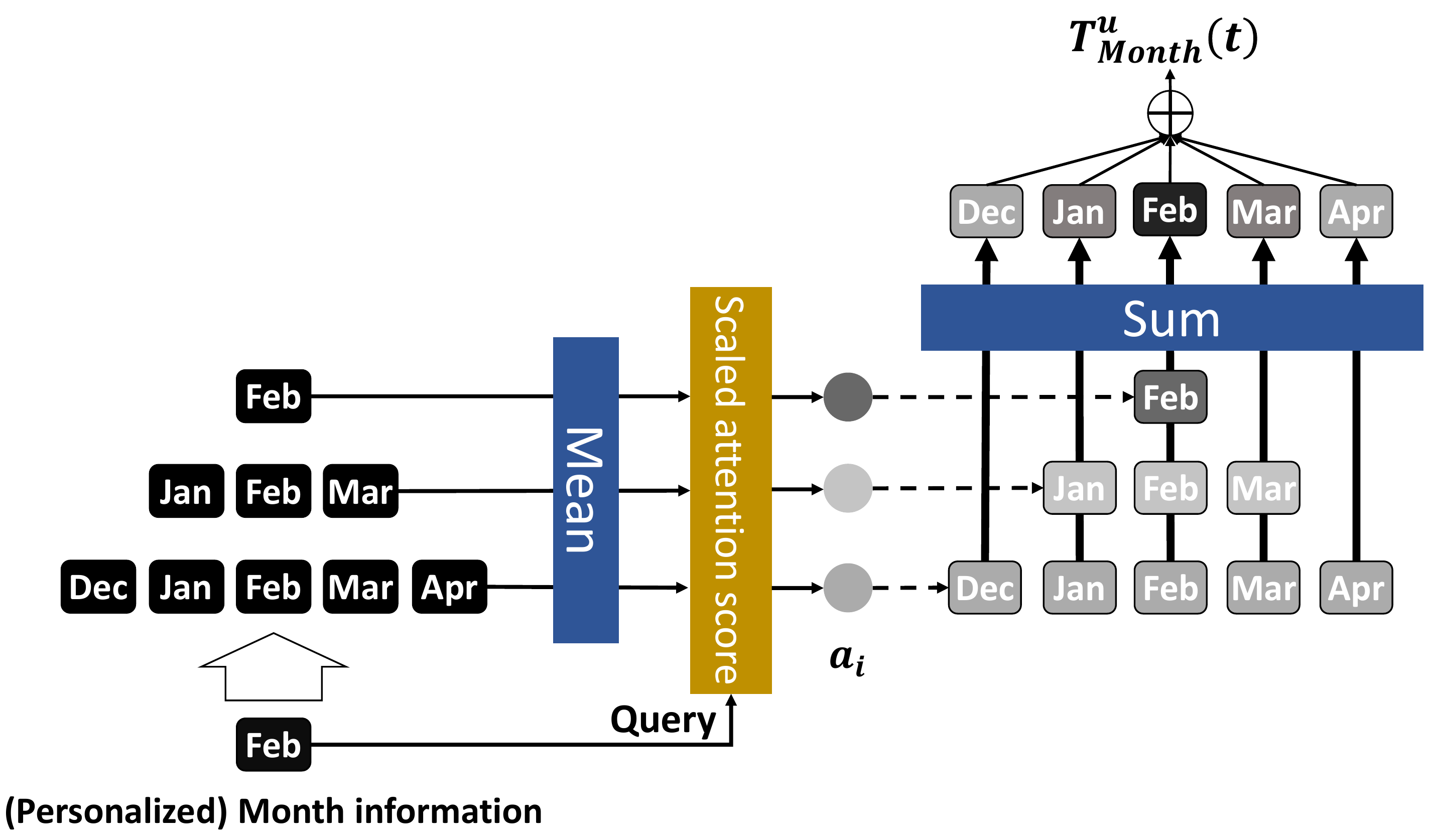}
    \caption{Visualization for the mechanism of gradual attention when the target month is February and $r_M$=2.}
    \label{fig:gradual}
\end{figure}
 
  We implement this idea with a novel attention module that we propose, called \textit{gradual attention} (Fig. \ref{fig:gradual}). Gradual attention adaptively combines the target time slot and the surrounding time slots within a certain range to create a representation of a granularity of period information of $t$. 
  Formally, we obtain a final representation of a granularity of period information of $t$:
  \begin{equation*}
    \begin{split}
        {T'}^u_{Month}(t, 0) &= E^u_{Month}(t) \\
        {T'}^u_{Month}(t, j) &= {\frac{{\sum^{j}_{n=-j} E^u_{Month}\left(\boldsymbol{\Delta}_{Month}(t, n)\right)}}{{2j+1}}} \quad (\text{for } j \ne 0)
    \end{split}
  \end{equation*}
  \begin{equation}
    \begin{split}
        a_i = &{\frac{\exp\left(E^u_{Month}(t) \cdot {T'}^u_{Month}(t, i) \middle/ \sqrt{d}\right)} {\sum_{i'=0}^{r_{M}} \exp\left(E^u_{Month}(t) \cdot {T'}^u_{Month}(t, i') \middle/ \sqrt{d}\right)}}
    \end{split}
  \end{equation}
  \begin{equation*}
    \begin{split}
        T^u_{Month}(t) &= \sum_{i=0}^{r_{M}} a_i {T'}^u_{Month}(t, i) \qquad \qquad \qquad \qquad
    \end{split}
  \end{equation*}
  where ${T'}^u_{Month}(t, j) \in \mathbb{R}^{d \times 1}$ is a representation that summarizes the personalized month information of timestamp $t$ and the $2j$ month slots near $t$, $\boldsymbol{\Delta}_{Month}(t, n)$ is the timestamp $n$ months after the timestamp $t$, $a_i$ is the attention score with scaled dot-product attention \cite{sinusoidal} which means the similarity score between $E^u_{Month}(t)$ and ${T'}^u_{Month}(t, i)$, $\cdot$ is the dot product, $r_{M}$ is the maximum range for the surrounding time slots of month, and $T^u_{Month}(t) \in \mathbb{R}^{d \times 1}$ is the final representation of the month of timestamp $t$.
 
  Gradual attention involves more information of a surrounding time slot as it is closer to the target time slot. Also, the attention score depends on the user since the period information used in gradual attention is personalized for the user. As a result, gradual attention gives more weight on a surrounding time slot that is closer to the target time slot and is more important to the user. We expect that gradual attention can be used for any attribute where neighboring slots have similar characteristics, similarly to the periodic patterns in time.
 
  \subsubsection{Combining the Representations for Period Information}
  After applying the series of processes described earlier to each granularity of period information, we have four representations for period information, $T^u_{Month}(t)$, $T^u_{DayOfWeek}(t)$, $T^u_{Date}(t)$, and $T^u_{Hour}(t)$. We obtain the final representation of $t$ by including each of the representations for period information as much as it is important to the user. That is, the final representation of timestamp $t$ is obtained as follows:
  \begin{equation}
      \begin{split}
          C &= \{Month, DayOfWeek, Date, Hour\}\\
          b_i &= \sigma\left( WU_u \cdot T^u_i(t)\right) \text{ for } i \in C\\
          T^u(t) &= \sum_{i \in C} b_i T^u_i (t) 
      \end{split}
  \end{equation}
  where $C$ is the set of the kinds of granularity of period, $b_i$ is the attention score for $i \in C$, $\sigma$ is the sigmoid function, $W \in \mathbb{R}^{d \times d}$ is the weight matrix for query of the multiplicative attention \cite{multiplicative}, and $T^u(t) \in \mathbb{R}^{d \times 1}$ is the final representation of timestamp $t$. 

\subsection{Time-aware History Encoder (TAHE)}
 \label{time_aware_history_encoder}
 A user's preference for an item may evolve over time under the influence of significant recent events such as the user's recent interactions or events that create a temporal trend of items.
 To consider these temporal patterns of user preferences, TAHE combines the target user's recent interactions and the temporal positions of the interactions into a latent representation.
 
 \subsubsection{Summarizing the Recent Interactions}
  To effectively summarize the user's recent interactions, TAHE employs the output of MATE to calculate the similarity between the time of each recent interactions and the target time. 
  The key idea behind this concept is that among the recent interactions, those which are made at a time that has similar periodic patterns to the target time can be more important.
  
  In this regard, we propose a novel attention mechanism called \textit{time-based attention}, to independently incorporate each of the user's recent interactions based on its similarity with the target time. 
  Specifically, time-based attention calculates a similarity score between each of $l$ recent interactions of the target user and the target time, in terms of the output of MATE. Then the item information of each recent interaction is aggregated according to the similarity to create the final representation of the user's interaction history. Formally, we summarize the target user's interaction history:
  \begin{equation}
      \begin{split}
          \text{cos}(x_1, x_2) &= {\frac{x_1 \cdot x_2} {|x_1||x_2|}} \\
          c^u_j(t) &= {\frac{\text{cos}\left(T^u(t), T^u(t^u_{j})\right) + 1}{2}}\\
          H^u(t) &= \sum_{j=1}^l c^u_j(t) I_{i^u_{j}}(t)
      \end{split}
  \end{equation}
  where $\text{cos}(\cdot, \cdot)$ is the cosine similarity between two embeddings, $c^u_j(t) \in [0, 1]$ is the similarity score between the time of $j$-th recent interaction of $u$ and $t$, $t^u_j$ is the timestamp of the $j$-th recent interaction of $u$ before $t$, $i^u_j$ is the $j$-th recent item that user $u$ consumed before $t$, $I_{i^u_j}(t)$ is a representation of item $i^u_j$ at time $t$ which is described in detail in the next section, and $H^u(t) \in \mathbb{R}^{d \times 1}$ is the final representation of the user's interaction history. 
  
  Time-based attention incorporates information of the item of each recent interaction independently based on its similarity to the target time in terms of the output of MATE. Therefore, the latent representation of the user's interaction history is defined depending on the target time, which is suitable for tasks that expect timely recommendations depending on the input time.

 \subsubsection{Temporal Position of Interactions}
  A user's interest in an item is susceptible to the temporal trend of the item, such as vogue or release of a film. 
  We argue that if the model can notice the temporal position of an interaction, it would be able to consider when an item is widely consumed by the users. 
  
  To this end, we employ positional encoding \cite{bert, sinusoidal}, which is a feature vector that expresses positional information of words in a sentence in natural language processing, to represent the temporal position of interactions. Among trainable positional embedding \cite{bert} and sinusoidal positional encoding \cite{sinusoidal}, we adopt sinusoidal positional encoding that does not have to be trained and thus is adaptable even to an unseen temporal position.
  As a result, the temporal position encoding $TE(t)$ of timestamp $t$ is defined as follows:
 \begin{equation}
     \begin{split}
         TE(t, j) &= 
         \begin{cases}
            \sin\left({\frac{t/3600} {10000^{j/d}}}\right), & \mbox{if }j\mbox{ is even} \\
            \cos\left({\frac{t/3600} {10000^{(j-1)/d}}}\right), & \mbox{if }j\mbox{ is odd}
         \end{cases}\\
         TE(t) &= \left[TE(t, 0), TE(t, 1), ..., TE(t, d-1)\right]^\top\\
         I_i(t) &= I_i + \alpha TE(t)
     \end{split}
 \end{equation}
 where $TE(t) \in \mathbb{R}^{d \times 1}$ is the temporal position encoding for timestamp $t$, $I_i \in \mathbb{R}^{d \times 1}$ is the item embedding for item $i$, $I_i(t) \in \mathbb{R}^{d \times 1}$ is the representation of item $i$ at timestamp $t$, and $\alpha$ is a trainable parameter for the weight of temporal position encoding. 
 
 We add the temporal position encoding not only to the item embedding of each of the user's recent interactions, but also to the target item embedding to consider the temporal position of the target time. We divide the timestamp by 3600 to convert it from seconds to hours, because seconds are too small to express the temporal position of an interaction. 
  
\subsection{Prediction}
 Finally, to calculate the interaction probability, TimelyRec incorporates the representation of the target timestamp $T^u(t)$, the representation of the target user's interaction history $H^u(t)$, user embedding $U_u$, and the representation of the target item at the target time $I_i(t)$. 
 TimelyRec takes account for the non-linear interactions of them via a multilayer perceptron (MLP) with ReLU \cite{relu} as activation function. That is, TimelyRec calculates the interaction probability $\hat{y}_{uit}$ as follows:
 \begin{equation}
 \begin{split}
    x_{uit} &= [U_u, I_i(t), T^u(t), H^u(t)]\\
    \hat{y}_{uit} &= \sigma\left(\text{MLP}(x_{uit})\right)
 \end{split}
 \end{equation}
 where $[\cdot,\cdot]$ is the concatenation, and $\sigma$ is the sigmoid function. 
 
\subsection{Model Optimization}
 Lastly, we optimize the model with the loss of the predictions $L$ using binary cross entropy function:
\begin{equation}
    L = - {\frac{1}{N}} \sum_{(u,i,t) \in D} \left(y_{uit} \log \hat{y}_{uit} + (1-y_{uit})\log (1-\hat{y}_{uit})\right)
\end{equation}
  where $N$ is the number of data in the training set, $D$ is the training set, and $y_{uit} \in \{0, 1\}$ is the ground-truth label of the interaction $(u, i, t)$ which 1 means a positive interaction, and 0 means a negative interaction.

\section{Experiment}
 In this section, we evaluate TimelyRec and the state-of-the-art time-aware methods on two time-related tasks: item recommendation, and item-timing recommendation. 

\subsection{Experimental Scenarios}
 \subsubsection{Item recommendation}
\begin{figure}[t]
    \centering
    \includegraphics[width=.92\columnwidth, trim={0cm 0cm 0cm 0cm}] {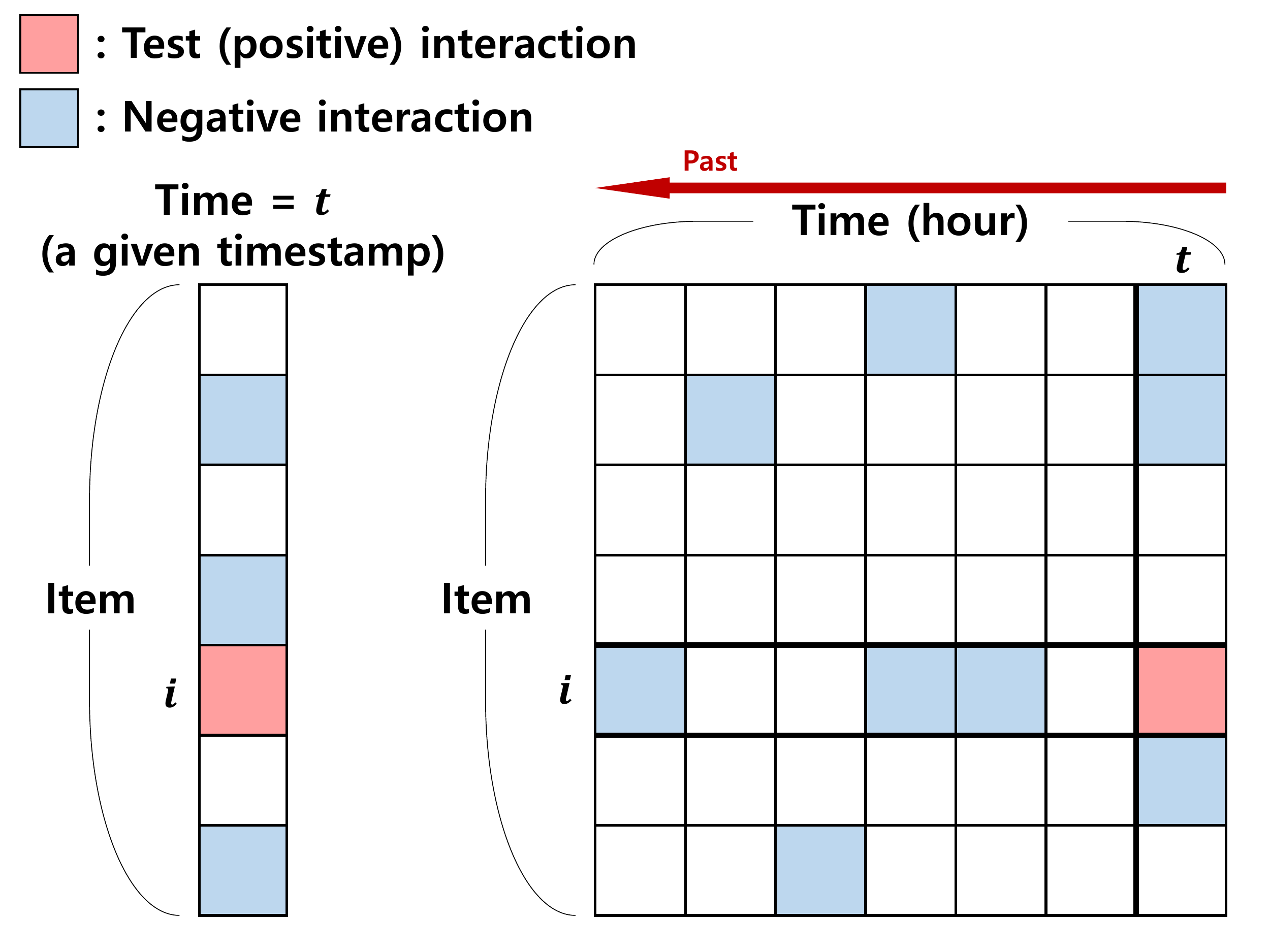}
    
    \caption{Evaluation scenario for item recommendation (left) and item-timing recommendation (right).}
    \label{fig:scenario}
\end{figure}

  Firstly, we evaluate the performance of the models on timely item recommendation in top-$k$ recommendation strategy (Fig. \ref{fig:scenario} (left)). We referred to the leave-one-out strategy \cite{strategy1, strategy2, strategy3, ndcg, ndcg2} which efficiently estimates the performance of a model on item recommendation. For each dataset, we choose the last interaction of each user as the test sample and the second last interaction as the validation sample, and utilize the remaining interactions for training. For a test interaction $(u, i, t)$ of user $u$, we build 100 random negative (i.e., unseen) interactions $(u, i^-, t)$ for the given time $t$, where $i^-$ means a negative item that the user has not consumed before. Then we rank the test interaction among the negative interactions. 
  
 \subsubsection{Item-timing recommendation}
  We conduct another experiment in addition to the item recommendation, which evaluates the performance of the models on item-timing recommendation in top-$k$ recommendation strategy (Fig. \ref{fig:scenario} (right)). The concept of our scenario for item-timing recommendation is not much different from the scenario for item recommendation, but we create three sets of negative interactions for a test interaction $(u, i, t)$ of user $u$, each of which consists of 100 random negative interactions. The first set consists of $(u, i^-, t)$ for the given time $t$, which is the same as the negative interactions in item recommendation. The second set contains $(u, i, t^-)$ for the given item $i$, where $t^-$ means a sampled negative timestamp, that is, a timestamp when user $u$ has never interacted with item $i$. Note that a negative timestamp is a past time no more than six months from the positive timestamp, and all the negative timestamps are sampled to be at least an hour apart. Lastly, the third set includes $(u, i^-, t^-)$. Then, we rank the test interaction among these 300 negative interactions. As the model should be able to predict not only an item that are suitable for the time but also a time that are suitable for the item, this evaluation scenario can evaluate the ability of a model to learn user preferences for items over time.
  
 To train the models to learn temporal patterns of user preferences, we constructed the training data with the positive interactions for training and three kinds of negative interactions. Similarly to the item-timing recommendation strategy, the three kinds of negative interactions are comprised of $(u, i^-, t)$, $(u, i, t^-)$, and $(u, i^-, t^-)$, respectively. The negative timestamps for a user are sampled between the user's first interaction time and the user's last interaction time. Also each negative interaction is sampled so that the model can distinguish it from the positive interactions. Lastly, the three negative interactions for each positive interaction are sampled again for each learning epoch.

\subsection{Experimental Settings}
\subsubsection{Datasets}
 We conducted our experiments on three public datasets: \textit{MovieLens}\footnote{https://grouplens.org/datasets/movielens/1m/} \cite{movielens}, \textit{Amazon Books}\footnote{http://jmcauley.ucsd.edu/data/amazon/} \cite{amazon}, and \textit{LastFM}\footnote{https://www.last.fm/} \cite{lastfm}. Table \ref{table:datasets} summarizes the statistics of each dataset. 
 
 \textbf{MovieLens.} MovieLens has 35-month user-movie interaction data. The users in this dataset had consumed at least 20 items. We retained only users who had accumulated interaction data for more than six months so that the model can learn enough the temporal patterns of a user's preference.
 
 \textbf{Amazon Books.} Amazon Books is the largest of Amazon datasets, which has 219-month user-book purchase data. We retained only users who have at least two years of interaction data, and eliminated users and items that have less than 15 interactions. Because Amazon Books does not contain the hour information, the negative interactions for this dataset were sampled to be at least a day apart instead of an hour. 
 
 \textbf{LastFM.} LastFM has 104-month user-artist music listening interaction data. Only users who have at least three years of interactions are retained, and we filtered out users and items that have less than 30 distinct interactions. Unlike the other datasets, most of the users had interacted with the same item several times in this dataset. Therefore, as the test sample of a user, we choose the last one of the first interactions with items which the user consumed at least three times. The validation sample is the second last among them. Interactions after the validation sample are not included in the training data. 
 
\begin{table}[t]
\centering
\caption{Statistics of datasets.}
\begin{tabular}{lccc}
\hline
Dataset     & \#users & \#items & \#interactions \\ \hline
MovieLens   & 866     & 3,548   & 289,170   \\
Amazon Books& 20,877  & 31,632  & 1,196,224 \\
LastFM      & 238     & 4,994   & 6,390,568 \\ \hline 
\end{tabular}
\label{table:datasets}
\end{table}

\begin{table*}
\fontsize{8.0}{3}\selectfont
\setlength{\tabcolsep}{1.2 pt}
\centering
\caption{Overall performance evaluation on item recommendation. $Imp.$ is the improvement of the performance from TimelyRec compared to the best performance among the others. The best results are highlighted in boldface, and the second best results are underlined.}
\def\arraystretch{3.5}
\begin{tabular}{c|c|ccccccccccccc|c|c} 
\hline
Dataset                                                                  & Metric  & LMF             & PITF$_{M}$  & PITF$_{W}$  & PITF$_{D}$  & PITF$_{H}$  & PITF$_{all}$  & TimeLSTM & SASRec & CoNCARS & NTF    & SLi-Rec         & GRec   & TiSASRec       & TimelyRec        & $Imp.(\%)$   \\ 
\hline
\multirow{5}{*}{MovieLens}                                               
& HR@1    & 0.1316          & 0.1127      & 0.1330      & 0.1164      & 0.1081      & 0.1328        & 0.1566   & 0.1351 & 0.1145  & 0.1247 & 0.1582          & 0.1525 & \underline{0.1619} & \textbf{0.1812}  & 11.92\%      \\
& HR@5    & 0.3760          & 0.3575      & 0.3780      & 0.3508      & 0.3536      & 0.3765        & 0.4215   & 0.3799 & 0.3378  & 0.3471 & \underline{0.4266}  & 0.3934 & 0.4194         & \textbf{0.4592}  & 7.65\%       \\
& NDCG@5  & 0.2563          & 0.2370      & 0.2592      & 0.2348      & 0.2318      & 0.2568        & 0.2932   & 0.2583 & 0.2273  & 0.2372 & 0.2956          & 0.2764 & \underline{0.2972} & \textbf{0.3238}  & 8.93\%       \\
& HR@10   & 0.5430          & 0.5356      & 0.5446      & 0.5483      & 0.5326      & 0.5564        & 0.5656   & 0.5413 & 0.5113  & 0.5018 & \underline{0.5681}  & 0.5396 & 0.5598         & \textbf{0.6039}  & 6.30\%       \\
& NDCG@10 & 0.3100          & 0.2943      & 0.3131      & 0.2983      & 0.2896      & 0.3147        & 0.3400   & 0.3086 & 0.2833  & 0.2870 & \underline{0.3416}  & 0.3197 & 0.3399         & \textbf{0.3696}  & 8.18\%       \\ 
\hline
\multirow{5}{*}{\begin{tabular}[c]{@{}c@{}}Amazon\\ Books \end{tabular}} & HR@1    & 0.1532          & 0.1881      & 0.1309      & 0.1142      & -           & 0.1768        & 0.2711   & 0.2420 & 0.1125  & 0.3196 & 0.3097          & 0.2600 & \underline{0.3208} & \textbf{0.3619}  & 12.81\%      \\
                                                                         & HR@5    & 0.4559          & 0.4571      & 0.4095      & 0.3755      & -           & 0.4774        & 0.5839   & 0.5231 & 0.3453  & 0.6212 & 0.6311          & 0.5663 & \underline{0.6361} & \textbf{0.6682}  & 5.05\%       \\
                                                                         & NDCG@5  & 0.3061          & 0.3281      & 0.2713      & 0.2464      & -           & 0.3309        & 0.4357   & 0.3902 & 0.2287  & 0.4801 & 0.4787          & 0.4199 & \underline{0.4841} & \textbf{0.5248}  & 8.42\%       \\
                                                                         & HR@10   & 0.6322          & 0.6066      & 0.5769      & 0.5421      & -           & 0.6515        & 0.7134   & 0.6488 & 0.5195  & 0.7364 & \underline{0.7515}  & 0.7077 & 0.7508         & \textbf{0.7796}  & 3.74\%       \\
                                                                         & NDCG@10 & 0.3658          & 0.3743      & 0.3256      & 0.3014      & -           & 0.3853        & 0.4778   & 0.4317 & 0.2829  & 0.5188 & 0.5203          & 0.4645 & \underline{0.5220} & \textbf{0.5599}  & 7.26\%       \\ 
\hline
\multirow{5}{*}{LastFM}                                                  & HR@1    & 0.0697          & 0.0722      & 0.0580      & 0.0521      & 0.0857      & 0.0538        & 0.0949   & 0.0563 & 0.1353  & 0.0857 & 0.1311          & 0.0974 & \underline{0.1404} & \textbf{0.2009}  & 43.08\%      \\
                                                                         & HR@5    & 0.2252          & 0.2143      & 0.1882      & 0.1715      & 0.2227      & 0.2126        & 0.2496   & 0.1841 & 0.2681  & 0.2118 & 0.2614          & 0.2521 & \underline{0.2773} & \textbf{0.3386}  & 22.11\%      \\
                                                                         & NDCG@5  & 0.1439          & 0.1460      & 0.1294      & 0.1245      & 0.1441      & 0.1352        & 0.1710   & 0.1268 & 0.2013  & 0.1526 & 0.1949          & 0.1723 & \underline{0.2106} & \textbf{0.2680}  & 27.30\%      \\
                                                                         & HR@10   & 0.3967  & 0.3176      & 0.3369      & 0.3084      & 0.3655      & 0.3126        & 0.3563   & 0.3294 & 0.3958  & 0.3092 & 0.3790          & 0.3596 & \underline{0.4110} & \textbf{0.4714}  & 14.71\%      \\
                                                                         & NDCG@10 & 0.2096          & 0.1772      & 0.1771      & 0.1625      & 0.1872      & 0.1670        & 0.2014   & 0.1744 & 0.2416  & 0.1810 & 0.2358          & 0.2020 & \underline{0.2477} & \textbf{0.3125}  & 26.16\%      \\
\hline
\end{tabular}
\label{table:item}
\end{table*}

\begin{table}
\fontsize{7.2}{2.7}\selectfont
\setlength{\tabcolsep}{1.2 pt}
\centering
\caption{Overall performance evaluation on item-timing recommendation.}
\def\arraystretch{3.5}
\begin{tabular}{c|c|cccc|c|c} 
\hline
Dataset                                                                  & Metric  & PITF$_{all}$  & TimeLSTM & SLi-Rec & TiSASRec       & TimelyRec        & $Imp.(\%)$   \\ 
\hline
\multirow{5}{*}{MovieLens}                                               & HR@1    & 0.0360        & 0.1143   & 0.1046  & \underline{0.1169} & \textbf{0.1388}  & 18.79\%      \\
                                                                         & HR@5    & 0.1578        & 0.3028   & 0.3012  & \underline{0.3053} & \textbf{0.3450}  & 13.00\%      \\
                                                                         & NDCG@5  & 0.0926        & 0.2103   & 0.2046  & \underline{0.2122} & \textbf{0.2413}  & 13.73\%      \\
                                                                         & HR@10   & 0.2524        & 0.4192   & 0.3986  & \underline{0.4263} & \textbf{0.4857}  & 13.93\%      \\
                                                                         & NDCG@10 & 0.1253        & 0.2460   & 0.2360  & \underline{0.2507} & \textbf{0.2858}  & 13.99\%      \\ 
\hline
\multirow{5}{*}{\begin{tabular}[c]{@{}c@{}}Amazon\\ Books \end{tabular}} & HR@1    & 0.0338        & 0.0848   & 0.1074  & \underline{0.1083} & \textbf{0.1195}  & 10.36\%      \\
                                                                         & HR@5    & 0.1264        & 0.1943   & 0.2351  & \underline{0.2462} & \textbf{0.3357}  & 36.37\%      \\
                                                                         & NDCG@5  & 0.0842        & 0.1399   & 0.1769  & \underline{0.1813} & \textbf{0.2092}  & 15.40\%      \\
                                                                         & HR@10   & 0.2017        & 0.2609   & 0.3039  & \underline{0.3203} & \textbf{0.4749}  & 48.26\%      \\
                                                                         & NDCG@10 & 0.1061        & 0.1612   & 0.1994  & \underline{0.2095} & \textbf{0.2577}  & 23.00\%      \\ 
\hline
\multirow{5}{*}{LastFM}                                                  & HR@1    & 0.0168        & 0.0353   & 0.0849  & \underline{0.0898} & \textbf{0.1345}  & 49.70\%      \\
                                                                         & HR@5    & 0.0395        & 0.1058   & 0.1589  & \underline{0.1647} & \textbf{0.2126}  & 29.13\%      \\
                                                                         & NDCG@5  & 0.0177        & 0.0672   & 0.1031  & \underline{0.1089} & \textbf{0.1713}  & 57.26\%      \\
                                                                         & HR@10   & 0.0697        & 0.1740   & 0.2210  & \underline{0.2288} & \textbf{0.2664}  & 16.40\%      \\
                                                                         & NDCG@10 & 0.0295        & 0.0921   & 0.1281  & \underline{0.1330} & \textbf{0.1838}  & 38.23\%      \\
\hline
\end{tabular}
\label{table:timing}
\end{table}

\subsubsection{Evaluation}
 We ranked the test interaction among the negative interactions and measured the performance of the models with two widely-used evaluation metrics for ranking \cite{strategy1, strategy2, strategy3, ndcg, ndcg2}, hit ratio (HR@$K$) and normalized discounted cumulative gain (NDCG@$K$). We use several values for $K$: 1, 5, and 10. NDCG@1 is omitted as it is the same as HR@1.
 
\subsubsection{Baselines}
 To demonstrate the effectiveness of TimelyRec on the tasks, we compare the performance of TimelyRec with several state-of-the-art baselines.
 \begin{itemize} 
     \item \textbf{LMF} \cite{lmf}: Logistic Matrix Factorization (LMF) is a Matrix Factorization for implicit feedback data which computes the interaction probability using the dot product of user embedding and item embedding.
     \item \textbf{PITF} \cite{pitf}: PITF computes the prediction score by adding the dot product of user-item, user-time, and item-time. We created several variants of PITF based on the granularity of period information that that model utilizes: month$(M)$, day of the week$(W)$, date$(D)$, hour$(H)$, and sum of all of their embeddings $(all)$.
     \item \textbf{TimeLSTM} \cite{timelstm}: TimeLSTM uses LSTM with time gates to consider the time interval between any two consecutive interactions. Among several versions of TimeLSTM, we used TimeLSTM3 which shows the best performance in \cite{timelstm}. 
     \item \textbf{SASRec} \cite{sasrec}: SASRec is a sequential recommender system without the time information, which adopts a self-attention network to capture the user's preference within a sequence.
     \item \textbf{CoNCARS} \cite{concars}: CoNCARS learns the temporal patterns of user and item behaviors with the time (hour) information with multiple convolutional layers. For Amazon Books, we used the day of the week instead of the hour. 
     \item \textbf{NTF} \cite{ntf}: NTF learns the non-linear interactions between user, item and the input time, which is expressed as the output of an LSTM which takes the last few months as inputs.
     \item \textbf{SLi-Rec} \cite{slirec}: SLi-Rec filters out unnecessary information in the user's short-term interaction history within TimeLSTM, and combines it with the user's long-term preference. We used the adaptive fusing method for SLi-Rec, which shows the best performance in \cite{slirec}. 
     \item \textbf{GRec} \cite{grec}: GRec is a state-of-the-art sequential recommender system without the time information, which leverages future data in a sequence as well for richer information in dilated convolutional neural networks.
     \item \textbf{TiSASRec} \cite{tisasrec}: TiSASRec is a state-of-the-art recommender system that utilizes the time information, which considers the time intervals between consecutive interactions in a user's interaction sequence in a self-attention network. 
 \end{itemize}

\subsubsection{Parameter Setting}
 We optimized all the models using Adam optimizer \cite{adam}. We tuned the hyperparameters by grid search with the validation HR@10 performance: learning rate $\eta$ $\in$ $\{0.01$, $0.001$, $0.0001\}$, dropout rate \cite{dropout} $p \in \{0.0, 0.1, 0.2, 0.3, 0.4, 0.5\}$ for each fully connected layer, width of the fully connected layers $width$ $\in$ $\{32$, $64$, $96$, $128$, $160\}$, and the number of the fully connected layers $depth \in \{1, 2, 3, 4, 5\}$. We added L2 regularization term for the methods without fully connected layers (i.e., LMF and PITF) with regularization parameter $\lambda \in \{0, 0.001, 0.01, 0.1\}$. 
 We set $l$ in TimelyRec and the number of time steps in NTF as 5. The dimension size $d$ is set to 32, and batch size is 256 in all the experiments. For TimeLSTM and SLi-Rec, the length of interaction history of each user is allowed up to 100, as in \cite{sasrec, slirec}. We set $r$, the maximum range for surrounding slots in gradual attention in TimelyRec, to about 20\% of the number of slots in each granularity of period: $r_M=2$, $r_W=1$, $r_D=6$, and $r_H=5$. 

\begin{figure*}[t]
    \centering
    \begin{subfigure}{0.39\columnwidth}
        \includegraphics[width=1.12\columnwidth, trim={0cm 0cm 0cm 0cm}, scale=0.2] {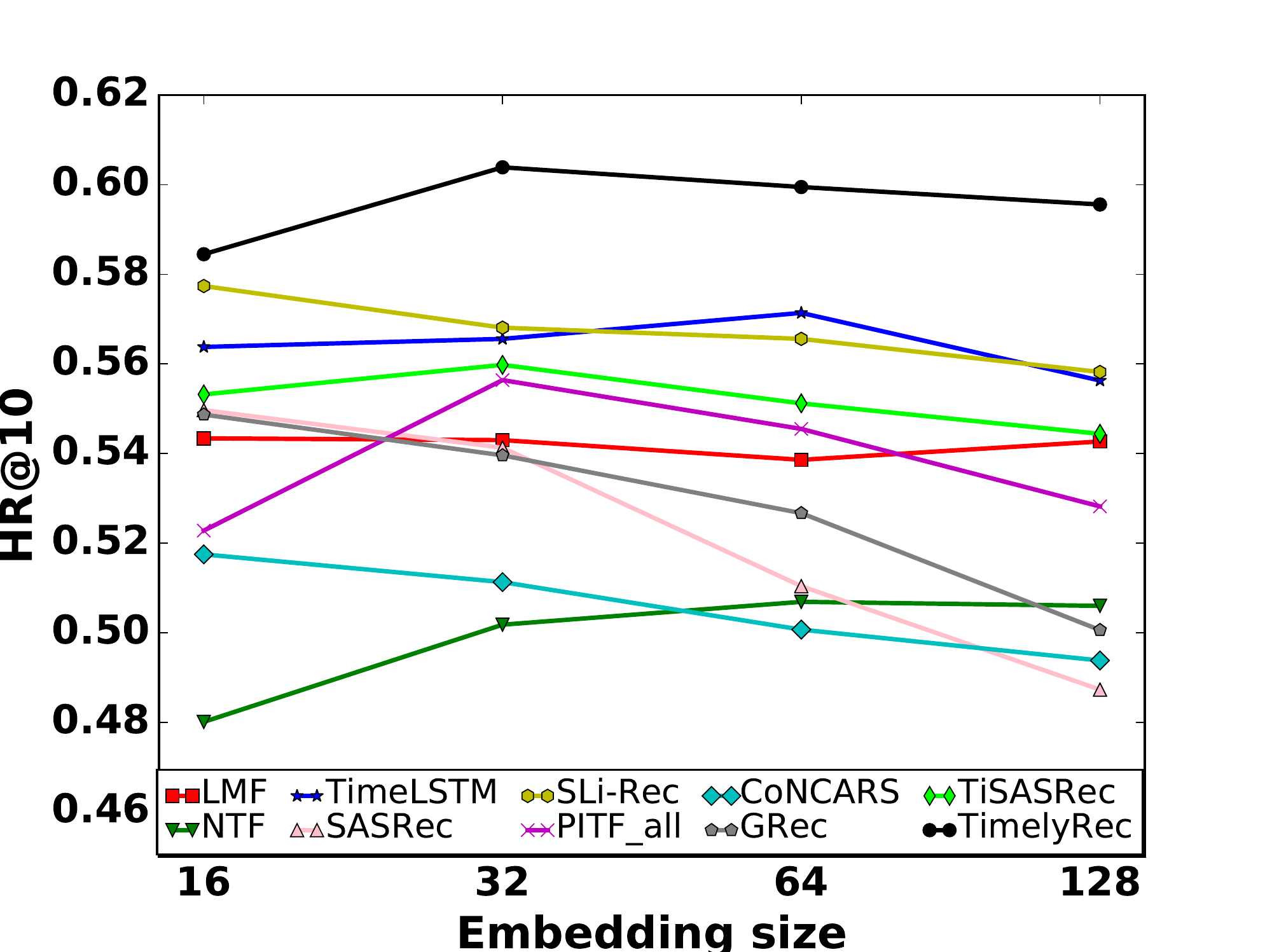}
        \subcaption{$d$ on MovieLens.}
    \end{subfigure}
    \begin{subfigure}{0.39\columnwidth}
        \includegraphics[width=1.12\columnwidth, trim={0cm 0cm 0cm 0cm}, scale=0.2] {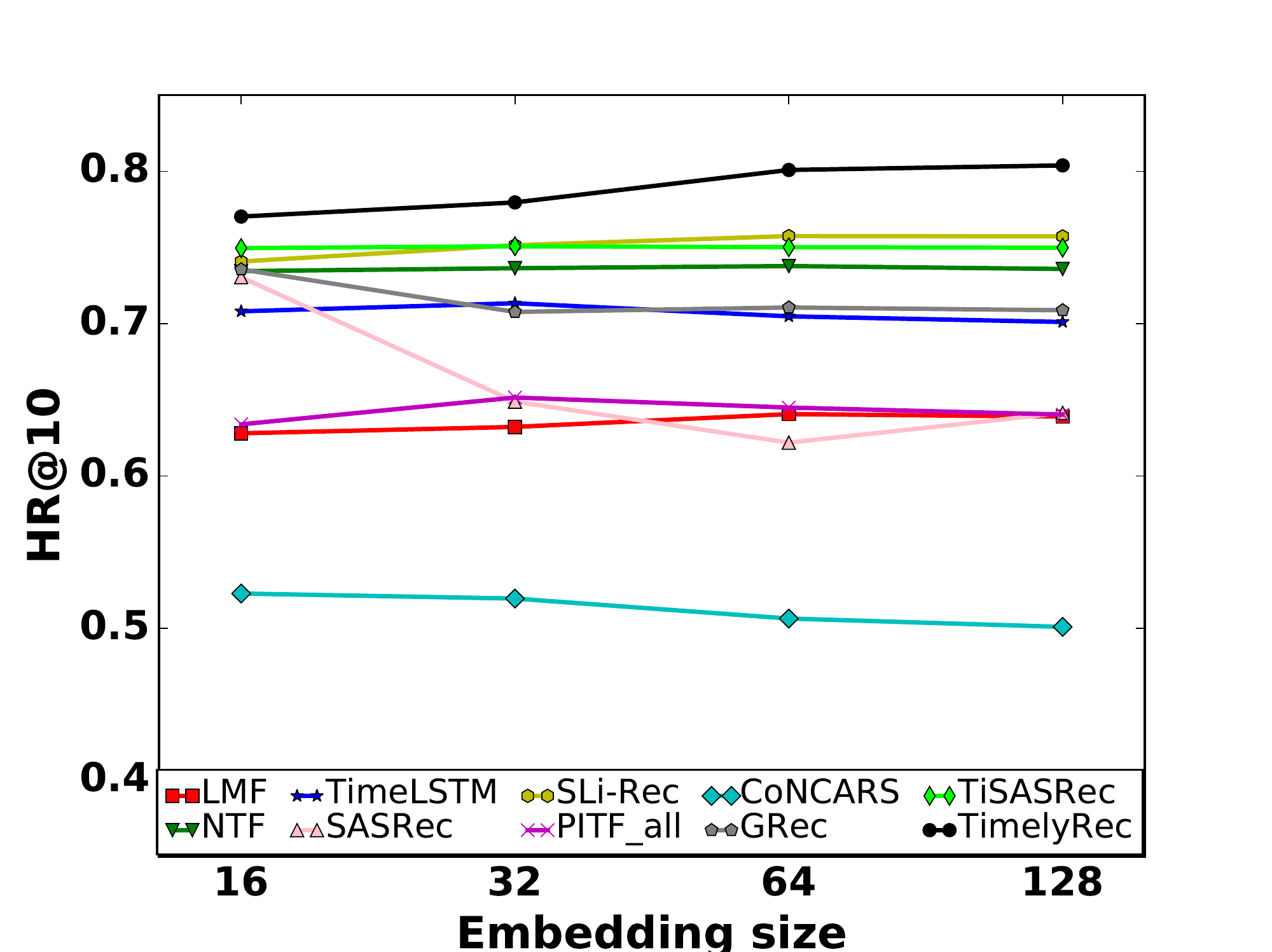}
        \subcaption{$d$ on Amazon Books.}
    \end{subfigure}
    \begin{subfigure}{0.39\columnwidth}
        \includegraphics[width=1.12\columnwidth, trim={0cm 0cm 0cm 0cm}, scale=0.2] {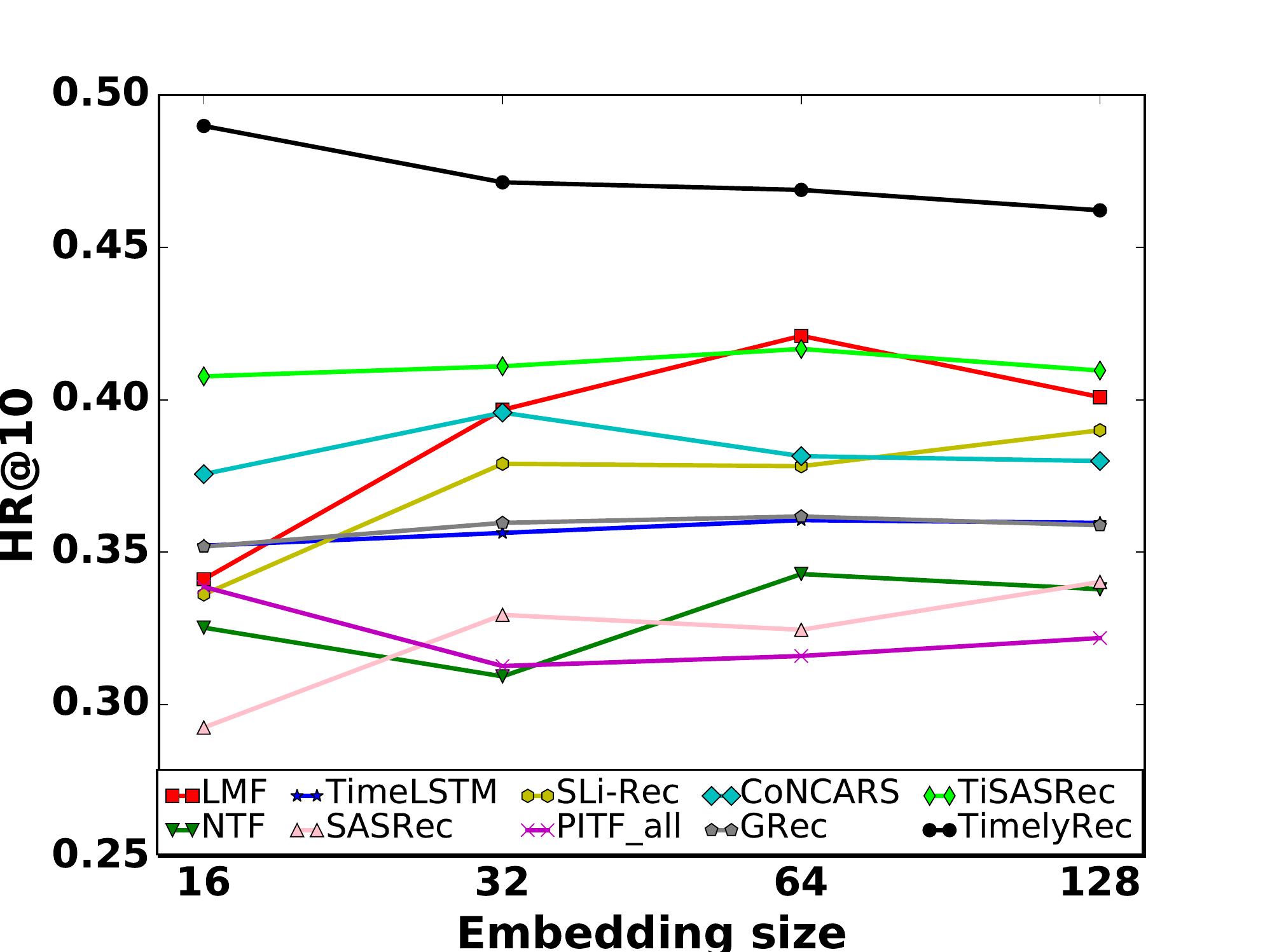}
        \subcaption{$d$ on LastFM.}
    \end{subfigure}
    \begin{subfigure}{0.39\columnwidth}
        \includegraphics[width=1.12\columnwidth, trim={0cm 0cm 0cm 0cm}, scale=0.2] {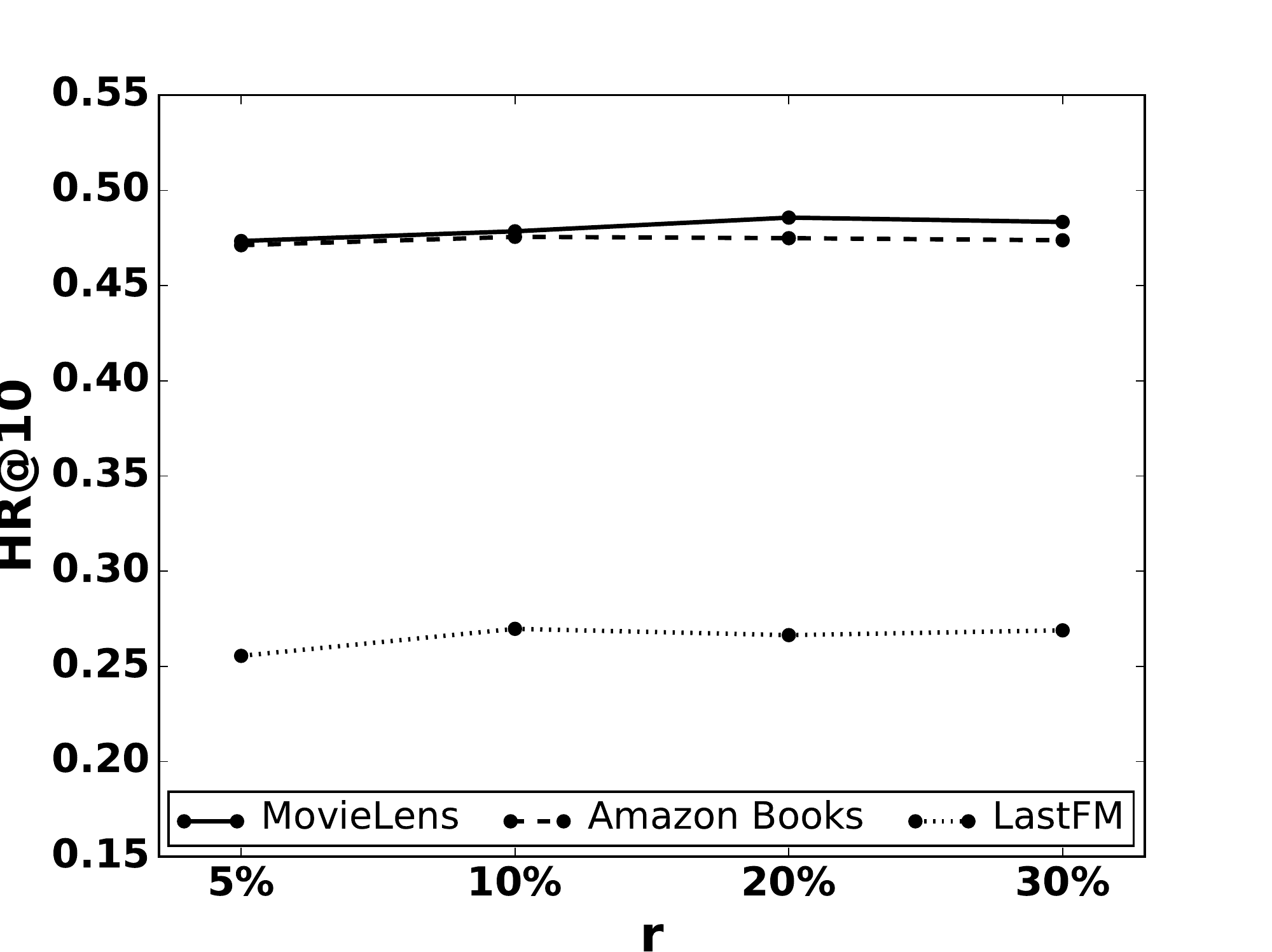}
        \subcaption{$r$ in TimelyRec.}
    \end{subfigure}
    \begin{subfigure}{0.39\columnwidth}
        \includegraphics[width=1.12\columnwidth, trim={0cm 0cm 0cm 0cm}, scale=0.2] {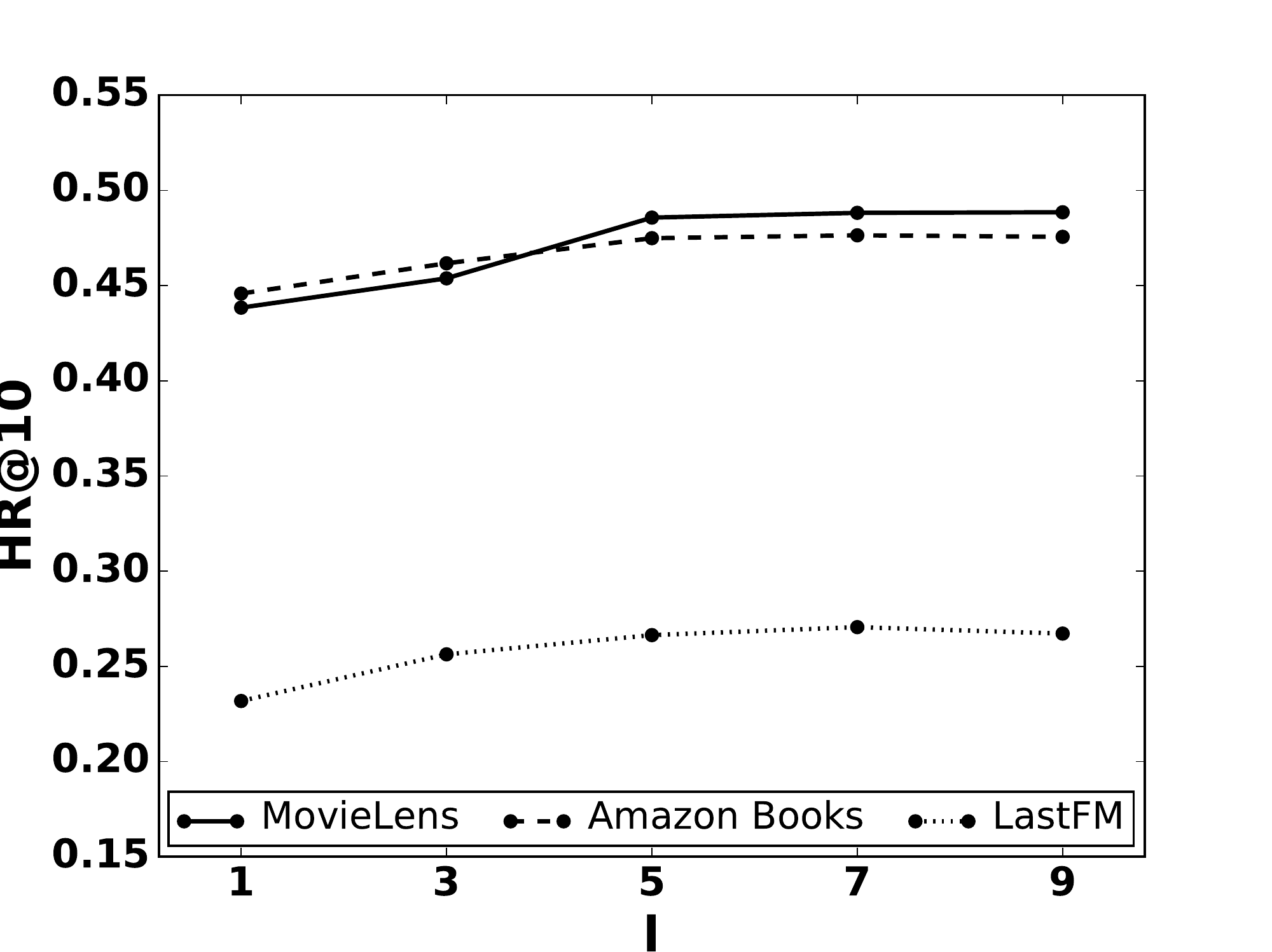}
        \subcaption{$l$ in TimelyRec.}
    \end{subfigure}
    \caption{Impact of each hyperparameter on the performance of the models (HR@10).}
    \label{fig:hyperparameter}
\end{figure*}

\subsection{Performance Comparison}
 We conducted the experiments on two tasks: item recommendation, and item-timing recommendation. For item-timing recommendation, we experimented only on models that can predict timing in hours. Each experiment was repeated five times, and we measured the performance on the test data at the epoch which shows the best validation performance (i.e., HR@10). Every result reported in each table is the average of the five results. 
 
 Table \ref{table:item} and Table \ref{table:timing} show the overall performance of the models on item recommendation and item-timing recommendation, respectively. We can draw some notable observations from the results. Firstly, our proposed TimelyRec outperforms all the baselines on both item recommendation and item-timing recommendation, for all the datasets. The performance improvements of TimelyRec compared to the best competitor are more evident in item-timing recommendation task. This result verifies the superiority of TimelyRec on learning the heterogeneous temporal patterns of user preferences, as the item-timing recommendation requires the ability to predict a time that is suitable for an item, in addition to an item that is suitable for a time. 
 
 We can also observe that, in Table \ref{table:item}, the baselines that use the time information show better performance in most cases than the methods that do not use the time information (i.e., MF, SASRec, and GRec). Also, TimelyRec which considers the heterogeneous characteristics of the recommending time shows the best performance among them. Therefore, it is important in the next item recommendation to consider the recommending time as well as the user's interaction history, especially to fully consider the heterogeneous characteristics of the recommending time personalized to the user. 

 Furthermore, in Table \ref{table:item}, the performance of PITF depends on which granularity of period is used, and the granularity that makes the best performance depends on the dataset. Also the naive integration of all the granularities of period (i.e., PITF$_{all}$) sometimes produces worse results than the other PITFs. These results verify our claim that the most significant granularity of period is different for each dataset, and each of them should not be integrated in a straightforward manner.

\begin{table}
\fontsize{7.5}{3}\selectfont
\setlength{\tabcolsep}{2 pt}
\def\arraystretch{3.3}
\centering
\centering
\caption{Result (HR@10) for ablation study of TimelyRec.}
\begin{tabular}{c|c|c|cc|cc|cc} 
\hline
\multicolumn{1}{c}{}    & \multicolumn{2}{c|}{}                    & \multicolumn{2}{c|}{MovieLens} & \multicolumn{2}{c|}{Amazon Books} & \multicolumn{2}{c}{LastFM}  \\ 
\hline
                        & \multicolumn{2}{c|}{Method}              & Item   & \begin{tabular}[c]{@{}c@{}}Item-\\ timing \end{tabular}                & Item   & \begin{tabular}[c]{@{}c@{}}Item-\\ timing \end{tabular}                   & Item   & \begin{tabular}[c]{@{}c@{}}Item-\\ timing \end{tabular}             \\ 
\hline
                        & \multicolumn{2}{c|}{TimelyRec}           & 0.6039 & 0.4857                & 0.7796 & 0.4749                   & 0.4714 & 0.2664             \\ 
\hline
\multirow{8}{*}{MATE}   & \multirow{4}{*}{a)} & No $Month$         & 0.5760 & 0.4702                & 0.7712 & 0.4693                   & 0.4471 & 0.2403             \\
                        &                     & No $DayOfWeek$     & 0.5646 & 0.4600                & 0.7697 & 0.4673                   & 0.4446 & 0.2361             \\
                        &                     & No $Date$          & 0.5580 & 0.4580                & 0.7722 & 0.4691                   & 0.4656 & 0.2622             \\
                        &                     & No $Hour$          & 0.5866 & 0.4774                & -      & -                        & 0.4227 & 0.2227             \\ 
\cline{2-9}
                        & b)                  & No personalization & 0.5716 & 0.4563                & 0.7599 & 0.4629                   & 0.4445 & 0.2437             \\ 
\cline{2-9}
                        & \multirow{2}{*}{c)} & No irregularity    & 0.5721 & 0.4536                & 0.7608 & 0.4672                   & 0.3900 & 0.2126             \\
                        &                     & No gradual att.   & 0.5848 & 0.4732                & 0.7679 & 0.4693                   & 0.4412 & 0.2479             \\ 
\cline{2-9}
                        & d)                  & Item query         & 0.5591 & 0.4501                & 0.7325 & 0.4185                   & 0.3579 & 0.2059             \\ 
\hline
\multirow{2}{*}{TAHE}   & e)                  & No time-based att. & 0.5781 & 0.4596                & 0.7628 & 0.4637                   & 0.4605 & 0.2496             \\ 
\cline{2-9}
                        & f)                  & No $TE(t)$         & 0.5760 & 0.4425                & 0.7708 & 0.4589                   & 0.4538 & 0.2437             \\ 
\hline
\multirow{2}{*}{Comb.~} & \multirow{2}{*}{g)} & No $T^u(t)$        & 0.5813 & 0.4474                & 0.7577 & 0.3971                   & 0.4026 & 0.2034             \\
                        &                     & No $H^u(t)$        & 0.5053 & 0.4164                & 0.6738 & 0.3419                   & 0.3916 & 0.1437             \\
\hline
\end{tabular}
\label{table:ablation}
\end{table}

\subsection{Ablation Study}
 We conducted an ablation study for the components of TimelyRec to verify the effectiveness of them. Table \ref{table:ablation} shows the overall HR@10 performance among all the variants of TimelyRec. We provide the analysis for each component in TimelyRec. 
 
 In MATE, a) each granularity of period information helps to improve the performance, which means all types of granularity provide meaningful information although each dataset has the most important type of granularity. b) Also our personalization improves the recommendation performance, as the same moment can have different meanings depending on the user. c) \emph{No Irregularity} is the result when the surrounding time slots are not considered. Capturing the irregularity of periodic pattern, which significantly improves the performance, plays a significant role in modeling the periodic patterns of user preference. Also our proposed gradual attention more effectively models the irregularity compared to a normal softmax attention across the surrounding time slots (i.e., \emph{No gradual att.}). Lastly, d) each granularity of period information should be included according to its importance to the user, whereas the item (i.e., \emph{Item query}) is not an appropriate criterion for the importance of each granularity.
 
 In TAHE, e) our proposed time-based attention with independent cosine similarity scores is more effective in aggregating the recent interactions of a user, compared to a normal softmax attention. f) Also our temporal position encoding $TE(t)$ helps to improve the recommendation performance by allowing the model to take account of the temporal position of the interactions. Finally, g) from the ablation study on when combining the representations, we can notice both the representation of the target timestamp and the target user's interaction history are important to the model's recommendation performance, as we expected. 

\begin{table*}[t]
\centering
\fontsize{8.2}{3}\selectfont
\setlength{\tabcolsep}{2 pt}
\caption{The attention scores in TimelyRec on an interaction example from MovieLens. 1st is the most recent interaction. }
\def\arraystretch{4.5}
\begin{tabular}{c|c|c|ccccccc|cccccccc|ccccc} 
\cline{1-11}\cline{18-23}
\multicolumn{1}{c}{}                   & \multicolumn{1}{c}{} &                      & \multicolumn{7}{c|}{Surrounding time slots}                                                                               &                & \multicolumn{1}{c}{} & \multicolumn{1}{c}{} & \multicolumn{1}{c}{} & \multicolumn{1}{c}{} & \multicolumn{1}{c}{} & \multicolumn{1}{c}{} & \multicolumn{1}{c|}{}            & \multicolumn{5}{c}{Recent interactions}                                                      \\ 
\cline{1-11}\cline{18-23}
Target time                            & \multicolumn{2}{c|}{Granularity}            
& Target        & 1st            & 2nd            & 3rd            & 4th            & 5th            & 6th                  & Importance     & \multicolumn{1}{c}{} &                      &                      & \multicolumn{1}{l}{} & \multicolumn{1}{c}{} &                      
& \multicolumn{1}{c|}{Granualrity} & \multicolumn{1}{c}{1st} & 2nd            & 3rd            &    4th         & 5th            \\ 
\cline{1-11}\cline{18-23}
\multirow{4}{*}{ \textbf{1045355829} } 
& $Month$                & \textbf{Feb}         & \textbf{1.0}  & \textbf{0.64}  & \textbf{0.22}  & \textbf{}      & \textbf{}      & \textbf{}      & \textbf{}            & \textbf{0.58}  &                      &                      &                      & \textbf{}            &                      &                     
& $Month$                            & \textbf{Feb}             & \textbf{Feb}   & \textbf{Jan}   & \textbf{Jan}   & \textbf{Jan}   \\
                                       
& $DayOfWeek$            & \textbf{Sun}         & \textbf{1.0}  & \textbf{0.36}  & \textbf{}      & \textbf{}      & \textbf{}      & \textbf{}      & \textbf{}            & \textbf{0.51}  &                      &                      &                      & \textbf{}            &                      &                      
& $DayOfWeek$                        & \textbf{Wed}             & \textbf{Sun}   & \textbf{Sun}   & \textbf{Mon}   & \textbf{Tue}   \\
                                       
& $Date$                 & \textbf{16}          & \textbf{1.0}  & \textbf{0.81}  & \textbf{0.67}  & \textbf{0.55}  & \textbf{0.40}  & \textbf{0.26}  & \textbf{0.13}        & \textbf{0.54}  &                      &                      &                      & \textbf{}            &                      &                      
& $Date$                             & \textbf{12}               & \textbf{2}    & \textbf{19}    & \textbf{13}     & \textbf{7}    \\
                                       
& $Hour$                 & \textbf{12 AM}        & \textbf{1.0}  & \textbf{0.69}  & \textbf{0.57}  & \textbf{0.44}  & \textbf{0.30}  & \textbf{0.15}  & \textbf{}            & \textbf{0.45}  &                      &                      &                      & \textbf{}            &                      &                      
& $Hour$                             & \textbf{4 AM}             & \textbf{11 PM}   & \textbf{5 PM}   & \textbf{3 AM}  & \textbf{3 AM}   \\ 
\cline{1-11}\cline{18-23}
\multicolumn{1}{c}{}                   & \multicolumn{1}{c}{} & \multicolumn{1}{c}{} &               &                &                &                &                &                & \multicolumn{1}{c}{} &                &                      &                      &                      & \textbf{}            &                      &                      
& \multicolumn{1}{c|}{Importance}  & \textbf{0.52}            & \textbf{0.48}  & \textbf{0.29}  & \textbf{0.41}  & \textbf{0.29}  \\
\cline{18-23}
\end{tabular}
\label{table:why}
\end{table*}

\begin{table*}[t]
\fontsize{6}{3}\selectfont
\setlength{\tabcolsep}{1 pt}
\centering
\caption{The similarity scores from time-based attention on an interaction example from MovieLens. The first instance is for the original interaction, the second is when another user replaces the target user, and the others are when another timestamp replaces the target time. 1st is the most recent interaction.}
\def\arraystretch{3}
\begin{tabular}{|c||c|c|ccccc|c|c|ccccc|c|c|ccccc|c|c|ccccc|} 
\hline
            & Target user & 15                 & \multicolumn{5}{c|}{Recent interactions}                                           & Target user & \textbf{525}       & \multicolumn{5}{c|}{Recent interactions}                                           & Target user & 15                    & \multicolumn{5}{c|}{Recent interactions}                                           & Target user & 15                    & \multicolumn{5}{c|}{Recent interactions}                                           \\ 
\hline
Granualrity 
& \multicolumn{2}{c|}{Target time} & 1st            & 2nd            & 3rd            & 4th            & 5th            & \multicolumn{2}{c|}{Target time} & 1st            & 2nd            & 3rd            & 4th            & 5th            & \multicolumn{2}{c|}{Target time}    & 1st            &2nd             & 3rd            & 4th            & 5th            & \multicolumn{2}{c|}{Target time}    & 1st            & 2nd            & 3rd            & 4th            & 5th            \\ 
\hline
$Month$       
& \multicolumn{2}{c|}{Dec}         & May            & Feb            & Feb            & Jan            & Dec            & \multicolumn{2}{c|}{Dec}         & May            & Feb            & Feb            & Jan            & Dec            & \multicolumn{2}{c|}{\textbf{Sep} }  & May            & Feb            & Feb            & Jan            & Dec            & \multicolumn{2}{c|}{\textbf{Feb} }  & May            & Feb            & Feb            & Jan            & Dec            \\
$DayOfWeek$   
& \multicolumn{2}{c|}{Fri}         &  Sun               &  Tue            & Fri            & Sun           & Mon        & \multicolumn{2}{c|}{Fri}         &  Sun               & Tue             & Fri            & Sun           & Mon        & \multicolumn{2}{c|}{\textbf{Thu} }  & Sun             &  Tue            & Fri            & Sun           & Mon           & \multicolumn{2}{c|}{\textbf{Sun} }  & Sun             & Tue             & Fri            & Sun           & Mon           \\
$Date$        
& \multicolumn{2}{c|}{6}           &  26                 & 19                  & 1              & 6        & 24        & \multicolumn{2}{c|}{6}           &  26                 &19                    & 1              & 6        & 24        & \multicolumn{2}{c|}{\textbf{26} }   & 26               & 19                   & 1              & 6        & 24           & \multicolumn{2}{c|}{\textbf{9} }    &  26              & 19                & 1              & 6           & 24           \\
$Hour$        
& \multicolumn{2}{c|}{8 AM}        & 4 AM           & 3 AM           & 6 AM           & 7 PM           & 5 AM           & \multicolumn{2}{c|}{8 AM}        &  4 AM          &  3 AM          & 6 AM           & 7 PM           & 5 AM           & \multicolumn{2}{c|}{\textbf{7 PM} } & 4 AM           &3 AM            & 6 AM           & 7 PM           & 5 AM           & \multicolumn{2}{c|}{\textbf{3 AM} } & 4 AM           &3 AM            & 6 AM           & 7 PM           & 5 AM           \\ 
\hline
Importance  
& \multicolumn{2}{c|}{}            & \textbf{0.48}  & \textbf{0.41}  & \textbf{0.39}  & \textbf{0.46}  & \textbf{0.43}  & \multicolumn{2}{c|}{}            & \textbf{0.35}  & \textbf{0.59}  & \textbf{0.41}  & \textbf{0.70}  & \textbf{0.55}  & \multicolumn{2}{c|}{}               & \textbf{0.40}  & \textbf{0.47}  & \textbf{0.38}  & \textbf{0.42}  & \textbf{0.28}  & \multicolumn{2}{c|}{}               & \textbf{0.29}  & \textbf{0.26}  & \textbf{0.28}  & \textbf{0.26}  & \textbf{0.34}  \\
\hline
\end{tabular}
\label{table:timebased}
\end{table*}

\begin{figure}
    \centering
    \begin{subfigure}{0.48\columnwidth}
        \centering
        \includegraphics[clip, width=1.06\linewidth, trim={1.5cm 0.7cm 2.5cm 1cm}] {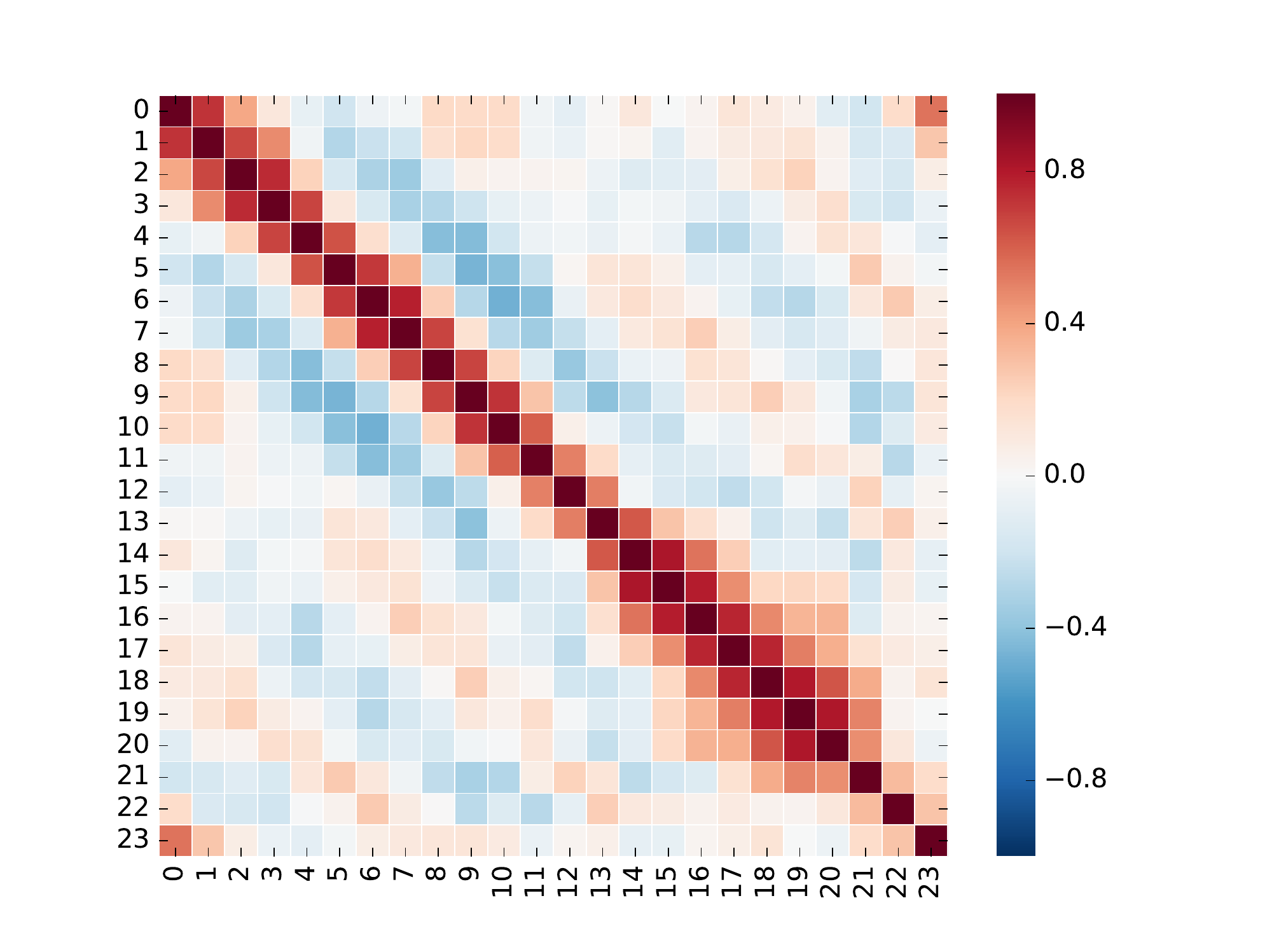}
        \caption{Hour}
    \end{subfigure}
    \hspace{0.1cm}
    \begin{subfigure}{0.48\columnwidth}
        \centering
        \includegraphics[clip, width=1.06\linewidth, trim={1.5cm 0.7cm 2.5cm 1cm}] {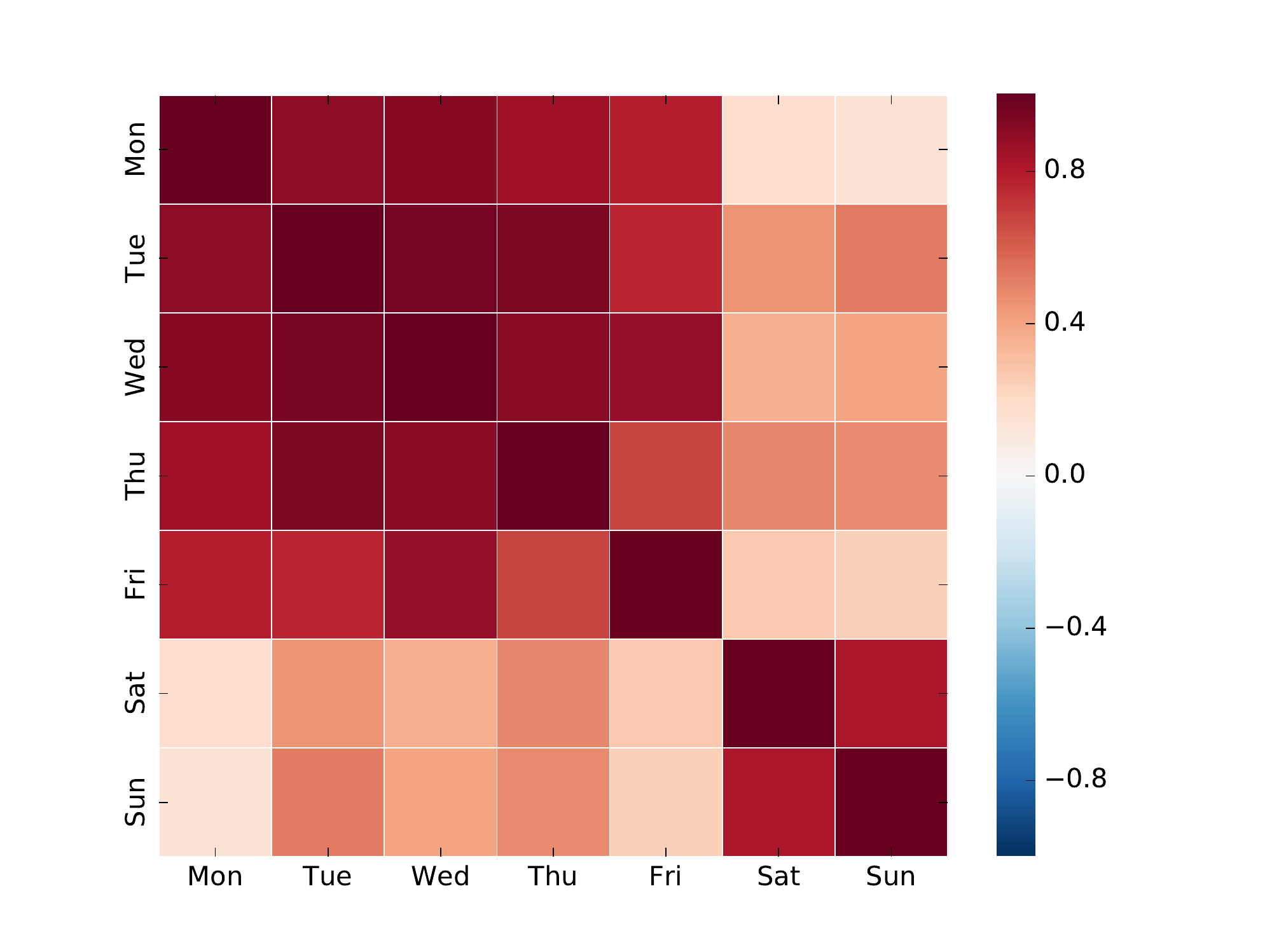}
        \caption{Day of the week}
    \end{subfigure} 
    \par \bigskip
    \begin{subfigure}{0.48\columnwidth}
        \centering
        \includegraphics[clip, width=1.06\linewidth, trim={1.5cm 0.7cm 2.5cm 1cm}] {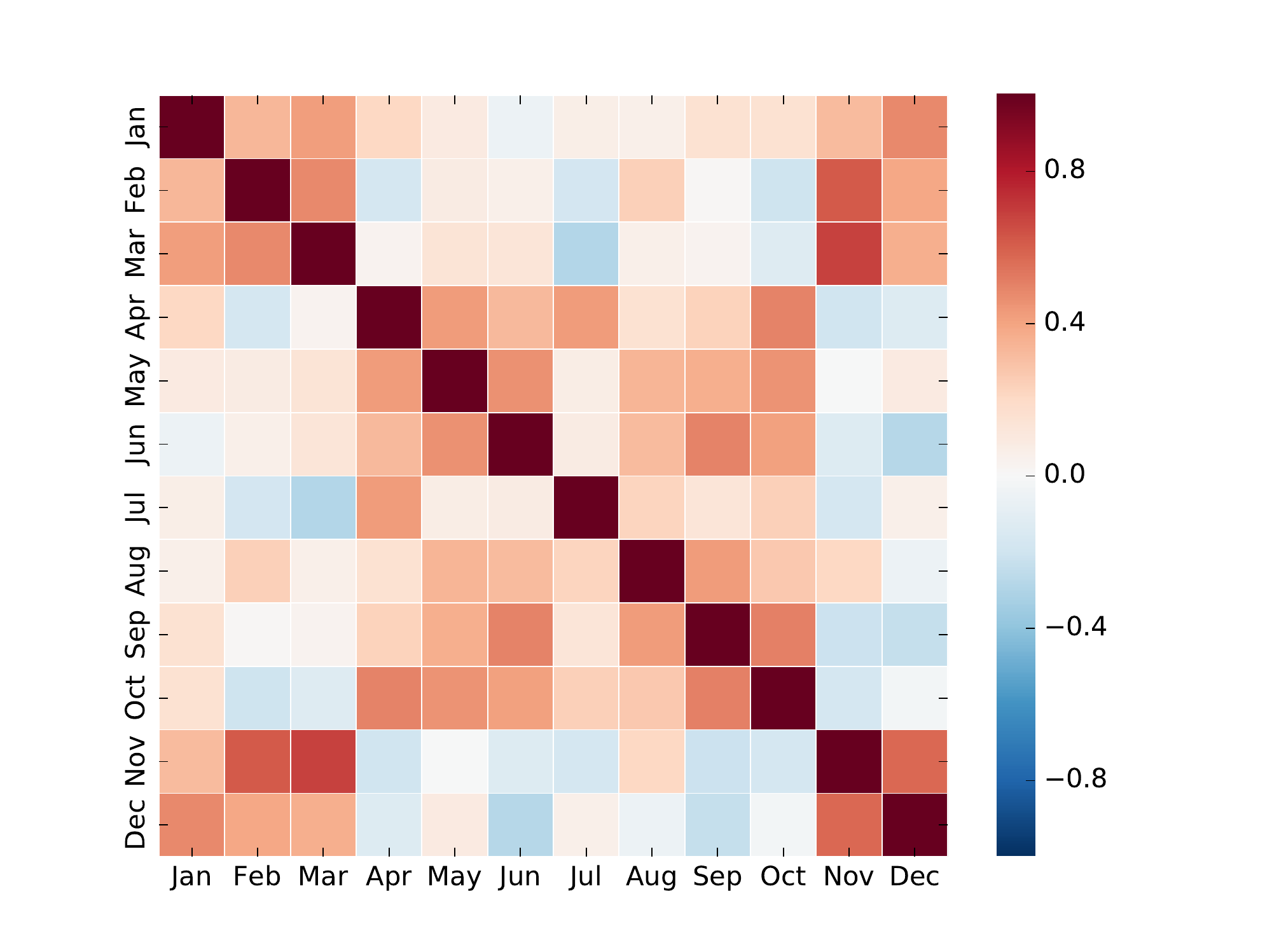}
        \caption{Month}
    \end{subfigure}
    \hspace{0.1cm}
    \begin{subfigure}{0.48\columnwidth}
        \centering
        \includegraphics[clip, width=1.06\linewidth, trim={1.5cm 0.7cm 2.5cm 1cm}] {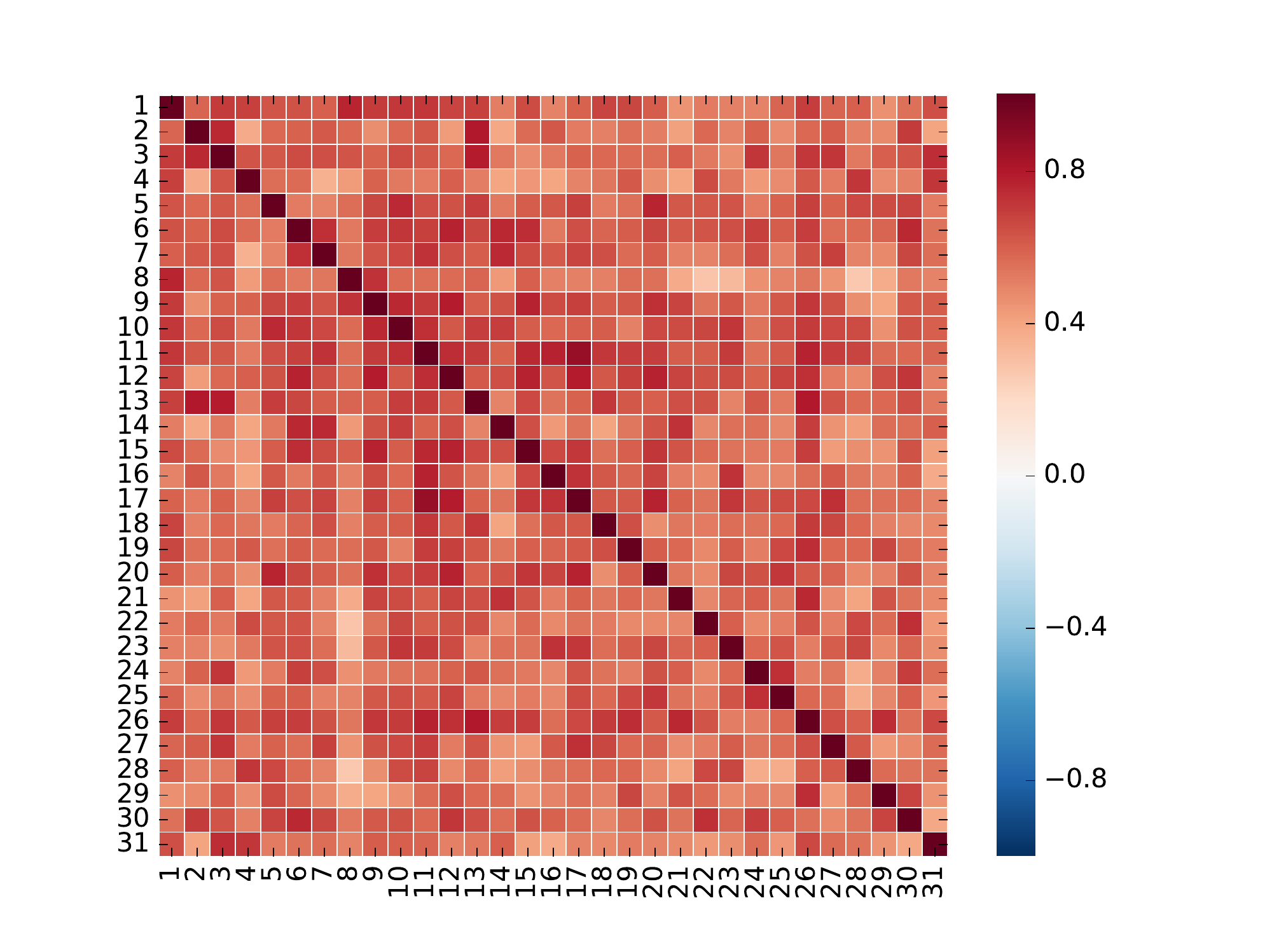}
        \caption{Date}
    \end{subfigure}

    \caption{Heatmap of cosine similarity between each pair of time slot embeddings for each granularity of period on LastFM.}
    \label{fig:heatmap}
\end{figure} 

\subsection{Hyperparameter Study}
 We analyse the effect of the hyperparameter selection in the methods: dimension size $d \in \{16, 32, 64, 128\}$, the maximum range for surrounding time slots in gradual attention $r \in \{5\%, 10\%, 20\%, 30\%\}$, and the number of the user's recent interactions $l \in \{1, 3, 5, 7, 9\}$. Fig. \ref{fig:hyperparameter} shows the trend of HR@10 performance as the value of each hyperparameter varies. 
 
 Fig. \ref{fig:hyperparameter}(a), (b), and (c) shows the impact of the embedding size on the performance on item recommendation for each dataset. The performance of TimelyRec tends to decrease when the embedding size is too large (i.e., 128) in the datasets except for Amazon Books. TimelyRec shows relatively stable performance on Amazon Books.

 Fig. \ref{fig:hyperparameter}(d) and (e) demonstrates the sensitivity of performance of TimelyRec on item-timing recommendation, based on each of its two hyperparameters. 
 $r$ is expressed as a percentage of the maximum range to the total number of slots in each granularity of period. The result shows that the performance of TimelyRec is stable when varying the number of surrounding slots in gradual attention. $l$ is the number of the user's recent interactions that TimelyRec incorporates. The performance of TimelyRec is degraded when the number of recent interactions in use is small, and becomes relatively stable when TimelyRec incorporates at least five recent interactions. 

\subsection{Qualitative Analysis on Behaviors of TimelyRec}
 In this section, we analyze the behaviors of TimelyRec. Table \ref{table:why} reports attention scores in TimelyRec when TimelyRec predicts the probability of a positive test interaction in MovieLens, which showed the highest prediction probability with TimelyRec. With this result, we can explain the reason why TimelyRec predicts the interaction as positive. In this example, we can judge that the user prefers to consume an item at a time similar to the target time (i.e., in February, around midnight, and on Sunday). Also the first, second, and fourth among the recent items were important to summarize the user's history, because the user interacted with them at a similar time to the target time. 

 We also analyze the ability of gradual attention that we propose to learn the periodic patterns of user preference. Fig. \ref{fig:heatmap} shows the heatmap for the cosine similarity between each pair of time slot embeddings before personalization on LastFM, for every granularity of period information. We can observe that, for hour in particular, the embeddings for neighboring time slots were learned to have similar values, because gradual attention trains the surrounding time slots at once. Also, there are some notable observations from the results: 1) The hour embeddings for day and evening (from 2 PM to 9 PM) were learned to be more similar to each other compared to those for night and morning. This result means there is no apparent change in user preference over time during the day and the evening. 2) When there is a difference of about 4 hours or 12 hours between the hour embeddings, they have the most opposing values. Therefore, we can guess that the user's preference is most conflicted between dawn (from 1 AM to 6 AM) and morning (from 5 AM to 10 AM) or day and night. 3) Although these tendencies are not as distinct as in the hour embeddings, the day of the week embeddings were learned to be divided into the weekend and the weekday. This result means there are differences between the temporal patterns on weekends and on weekdays. 4) In the month embeddings, we observe that the temporal patterns of user preference are grouped into months from April to October and months from November to March. 5) The date embeddings were trained to have no apparent tendency. Note that this is consistent with the experimental results in Table \ref{table:item} where PITF$_{D}$ shows the lowest performance among the versions of PITF, which means there is no particular pattern of user preference over date and therefore date is not a significant information in LastFM. 
 
 \balance
 Finally, we provide an analysis of the behaviors of time-based attention.
 Table \ref{table:timebased} reports the similarity scores from time-based attention with different target timestamps and target users. The positive test interaction used in this example is from MovieLens, which only TimelyRec estimates as top-1 among the baselines. We can observe that time-based attention produced the similarity scores for the recent interactions in a time-aware and personalized manner. Therefore, we argue that time-based attention can be effectively applied to the time-aware recommender systems, which provide personalized recommendations according to the user and the time. 

\section{Conclusion}
 In this paper, we first define various unique characteristics in the periodic pattern and the evolving pattern of user preference, which should be considered in time-aware recommender systems. Then we propose a novel recommender system for timely recommendations, called TimelyRec, which jointly considers all the unique characteristics of the heterogeneous temporal patterns of user preference, and two attention modules within the model. We also introduce an evaluation scenario for item-timing recommendation in top-$K$ recommendation. Experimental results and analyses on item recommendation and item-timing recommendation verify the superiority of TimelyRec and our proposed attention modules. 

\begin{acks}
This work was supported by the NRF grant funded by the MSIT (No. 2020R1A2B5B03097210), and the IITP grant funded by the MSIT (No. 2018-0-00584, 2019-0-01906).
\end{acks}

\bibliographystyle{ACM-Reference-Format}
\bibliography{main}










\end{document}